\documentclass[aps,10pt,twocolumn,amsmath,amssymb,showpacs,floatfix,longbibliography]{revtex4-2}
\usepackage{graphicx}
\usepackage{xcolor}
\usepackage{bm}
\usepackage[utf8]{inputenc}
\usepackage[T1]{fontenc}
\usepackage[%
  colorlinks=true,
  urlcolor=blue,
  linkcolor=blue,
  citecolor=blue]{hyperref}
\usepackage[all]{hypcap}
\usepackage{sidecap}
\sidecaptionvpos{figure}{t}
\usepackage[capitalise]{cleveref}
\graphicspath{{pictures/}}

\begin{document}

\title{Multifunctional steep-slope spintronic transistors with spin-gapless-semiconductor or spin-gapped-metal electrodes}


\author{Ersoy \c{S}a\c{s}{\i}o\u{g}lu$^1$}\email{ersoy.sasioglu@physik.uni-halle.de}
\author{Paul Bodewei$^{1}$}
\author{Nicki F. Hinsche$^1$}
\author{Ingrid Mertig$^{1}$}

\affiliation{$^{1}$Institute of Physics, Martin Luther University Halle-Wittenberg, 06120 Halle (Saale), 
Germany \\}

\date{\today}

\begin{abstract}

Spin-gapless semiconductors (SGSs) are emerging as a promising class of materials for spintronic 
applications, offering unique opportunities to realize novel functionalities beyond conventional 
electronics. In this work, we propose a novel concept of multifunctional spintronic field-effect 
transistors (FETs) by employing SGSs and/or spin-gapped metals (SGMs) as the source and drain 
electrodes. These devices function similarly to metal-semiconductor Schottky barrier FETs, where a
potential barrier forms between the SGS (SGM) electrode and the intrinsic semiconducting channel. 
However, unlike metal-semiconductor Schottky barrier FETs, the proposed devices leverage the
distinctive  spin-dependent transport properties of SGS and/or SGM electrodes to enable sub-60 
mV/dec switching, a significant improvement over the sub-threshold swing bottleneck of 60 mV/dec 
in conventional MOSFETs, facilitating low-voltage operation. Additionally, the incorporation of 
SGMs introduces the negative differential resistance (NDR) effect with an ultra-high 
peak-to-valley current ratio, further enhancing the device functionality. The proposed spintronic 
FETs exhibit a remarkable combination of features, including sub-60 mV/dec switching, non-local 
giant magnetoresistance (GMR) effect, and NDR effect, making them highly attractive for 
next-generation applications such as logic-in-memory computing and multivalued logic. These 
functionalities open avenues for exploring novel computing architectures beyond the constraints 
of the classic von-Neumann architecture, enabling more efficient and powerful information processing.
Two-dimensional (2D) nanomaterials emerge as a promising platform for realizing these 
multifunctional FETs.  In this work, we perform a comprehensive screening of the computational 
2D materials database to identify suitable SGS and SGM materials for the proposed devices. For 
device simulations, we select VS$_2$ as the SGS material. As a proof-of-concept, we employ a 
non-equilibrium Green's function method combined with density functional theory to simulate the 
transfer ($I_{\mathrm{D}}$-$V_{\mathrm{G}}$) and output ($I_{\mathrm{D}}$-$V_{\mathrm{D}}$)  
characteristics of a vertical VS$_2$/Ga$_2$O$_2$ heterojunction FET based on 2D type-II SGS VS$_2$. 
Our calculations predict a remarkably low sub-threshold swing of 20\,mV/dec, a high on/off ratio of 
10$^8$, and a significant non-local GMR effect demonstrate these devices' potential for low-power 
and high-performance applications.

\end{abstract}
\maketitle

%
%
%
%
%
%
%
%
%
%

\section{Introduction}

Modern charge-based electronics rely heavily on metal-oxide-semiconductor field-effect transistors 
(MOSFETs) as fundamental building blocks. Their remarkable scaling over the decades has revolutionized 
computing and communication technologies, enabling the development of increasingly powerful and compact 
devices. However, despite their widespread use and advancements, conventional MOSFETs face inherent 
limitations that challenge further progress in electronics scaling and energy efficiency. One such 
limitation is the 60 mV/dec sub-threshold swing (SS) imposed by thermionic emission, which has remained 
a persistent bottleneck \cite{taur2021fundamentals,lundstrom2002fundamentals}. This limitation significantly 
hinders low-voltage operation, a critical factor for achieving energy-efficient devices in modern electronics. 
Lowering the operating voltage not only reduces static power consumption but also minimizes leakage currents, 
addressing a fundamental concern for battery-powered electronics and portable devices where energy 
efficiency is paramount. Therefore, there is a pressing need to explore alternative transistor designs 
and materials that can overcome these limitations and pave the way for the next generation of energy-efficient 
electronic devices \cite{wong2015memory,ionescu2011tunnel,ferain2011multigate}.

To overcome the 60 mV/dec SS limit, researchers have explored alternative transistor designs 
categorized as steep slope transistors. These devices aim to achieve a sharper transition
between the on-state and off-state, leading to a lower SS value. Examples of such devices 
include tunnel FETs (TFETs), Dirac source FETs, and the recently proposed cold-metal source FETs 
\cite{ionescu2017beyond,qiu2018dirac,liu2020switching,tang2021steep,wang2023cold,wang2022road,yin2022computational}.
While these alternative designs offer lower SS compared to conventional MOSFETs, they often come with 
trade-offs. Tunnel FETs, for instance, rely on a tunneling mechanism for current flow, leading 
to significantly lower on-currents compared to MOSFETs. This drawback limits their applicability 
in many logic circuits where high on-currents are crucial for driving 
subsequent stages. In contrast, Dirac source and cold-metal source FETs generally exhibit much
higher on-state current densities, making them more attractive alternatives for practical applications. 
Steep-slope FETs with Dirac source and cold-metal source electrodes are an area of active research 
and are being explored as a potential solution to overcome the limitations of traditional MOSFETs, 
especially for low-power and high-speed applications. However, a key limitation of these emerging 
transistor proposals is their reliance solely on the charge degree of freedom. They do not exploit 
the potential benefits offered by the electron's spin, a valuable resource for future spintronic 
devices.

Spin gapless semiconductors (SGSs) have shown promising potential in magnetic tunnel junctions (MTJs), 
enabling functionalities beyond conventional MTJs based on Fe, Co, and CoFeB. While these traditional MTJs
offer high tunnel magnetoresistance (TMR) for memory applications, they lack current rectification, 
hindering their functionality as diodes for logic applications \cite{mathon2001theory,parkin2004giant,chappert2007emergence}. 
A recent proposal addressed this by introducing type-II SGSs and half-metallic magnets in MTJs, 
achieving both TMR and a re-programmable diode effect \cite{sasioglu2019proposal,aull2022ab}, which is 
experimentally demonstrated in MTJs based on Heusler compounds \cite{maji2022demonstration}. Inspired by 
this progress, we propose a novel class of spintronic FETs utilizing SGSs and/or spin-gapped metals (SGMs) 
as source and drain electrodes with an intrinsic semiconductor channel. This configuration overcomes the 
limitations of both conventional MOSFETs (restricted by the 60 mV/dec sub-threshold swing) and alternative 
steep-slope transistor designs. The key lies in exploiting the unique spin-dependent transport properties 
of SGS and SGM electrodes. The SGSs enable sub-60 mV/dec switching and introduce the non-local giant 
magnetoresistance (GMR) effect, absent in conventional charge-based FETs. Additionally, SGMs offer negative 
differential resistance  (NDR) with a high peak-to-valley current ratio. This combination opens doors for 
next-generation applications like logic-in-memory computing, where data processing and storage occur on the 
same chip \cite{linn2012beyond,you2014exploiting,gao2015implementation,zhou2015boolean,wang2017functionally,kim2019single}.

In this paper, we present a comprehensive conceptual framework for novel multifunctional spintronic 
FETs. Two-dimensional (2D) nanomaterials emerge as a promising platform for realizing these 
multifunctional FETs. These materials offer a promising pathway to overcome scaling challenges inherent 
to conventional FETs. Their atomically thin structure enables exceptional gate control and mitigates 
short-channel effects, essential for low-power, high-performance devices. Focusing on spintronic 
applications, we identify transition metal dichalcogenides and dihalides, such as VS$_2$ and ScI$_2$, 
as potential 2D candidates. Through a comprehensive screening of a computational 2D materials database, 
we select suitable SGS and SGM materials for our proposed devices. To demonstrate proof-of-concept, we 
employ ab initio quantum transport calculations to simulate a vertical VS$_2$/Ga$_2$O$_2$ 
heterojunction FET based on 2D type-II SGSs VS$_2$. Utilizing a non-equilibrium Green function method 
combined with density functional theory, our simulations predict remarkable device performance, including 
a subthreshold swing (SS) of only 20 mV/dec, a high on/off ratio of 10$^8$, and a substantial non-local 
giant magnetoresistance (GMR) effect. These findings underscore the potential of 2D SGSs for realizing 
our proposed multifunctional spintronic FETs, paving the way for a new generation of low-power, 
high-performance spintronic devices. The remainder of the paper is structured as follows: In Section II, 
we introduce the concepts of SGSs and  SGMs. Section III presents the 
design and operation principles of our proposed multifunctional spintronic FETs. The computational 
methodology employed in this study is detailed in Section IV. In Sections V.A and V.B, we present the 
results of our screening process for suitable 2D materials and the simulation results for a vertical 
VS$_2$/Ga$_2$O$_2$ heterojunction FET, respectively. Finally, Section VI summarizes our key findings 
and discusses their implications.

\begin{figure*}[t]
\centering
\includegraphics[width=0.9\textwidth]{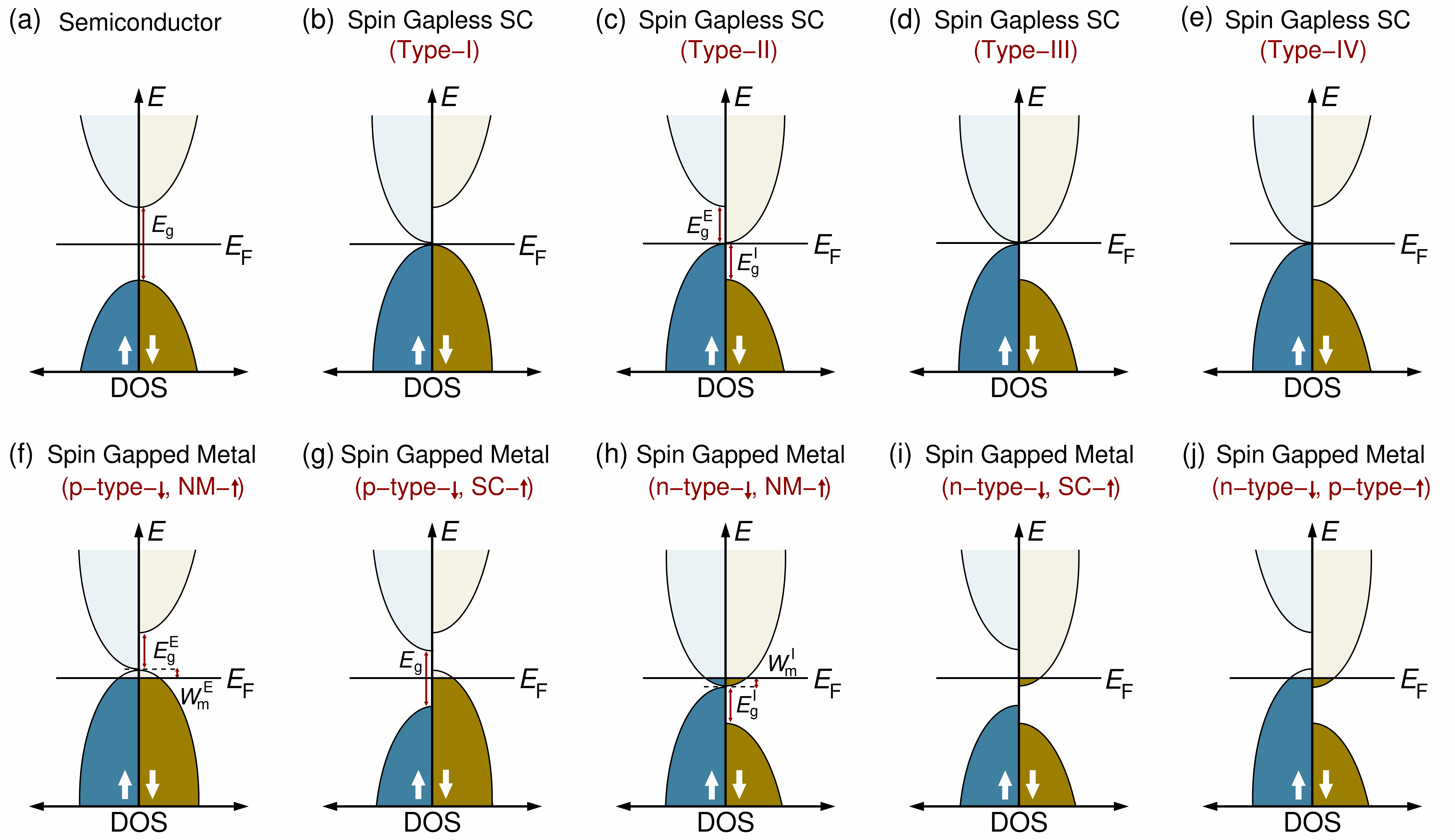}
\vspace{-0.2 cm}
\caption{Schematic representation of the density of states (DOS) of a semiconductor
(a), spin-gapless semiconductors (b-c), and spin-gapped metals (d-j). 
The arrows represent the two possible spin directions. The horizontal line 
depicts the Fermi level $E_\mathrm{F}$. NM stands for normal-metal and SC 
for semiconductor.}
\label{fig1}
\end{figure*}

\section{Spin gapless semiconductors and spin gapped metals}

SGSs have emerged as a promising class of materials for device applications due to their 
unique properties. The concept of SGSs was introduced by Wang in 2008 \cite{wang2008,wang2009colossal}. 
Using first-principles calculations, Wang predicted SGS behavior in Co-doped PbPdO$_2$ \cite{wang2008}. 
Since then, variousmaterials ranging from two-dimensional to three-dimensional structures 
have been theoretically predicted to exhibit SGS behavior \cite{wang2020spin}, with some 
confirmed experimentally \cite{ouardi2013realization}. SGSs occupy a space between magnetic 
semiconductors and half-metallic magnets (HMMs) \cite{de1983new}. Fig.~\ref{fig1} depicts 
a schematic density of states (DOS) for different types of SGSs. In type-I, type-III, and 
type-IV SGSs, the minority-spin band resembles that of HMMs, but the majority-spin band 
differs. The valence and conduction band edges touch at the Fermi energy, resulting in a 
zero-gap state. In contrast, type-II SGSs possess a unique band structure where a finite gap 
exists just above and below the Fermi level (E$_F$) for each spin channel. However, the 
conduction and valence band edges of the different spin channels touch. Importantly, SGSs 
exhibit either ferromagnetic or ferrimagnetic  behavior.

One key advantage of type-I, type-III, and type-IV SGSs is that exciting electrons from 
the valence to the conduction band require no energy, and the excited electrons or holes 
can be 100\% spin-polarized. Similarly, no energy is needed for spin-flipped Stoner excitations 
in type-II SGSs. Notably, the mobility of charge carriers in SGSs is generally higher than 
in conventional semiconductors, making them attractive for nanoelectronic applications. 
Furthermore, the unique spin-dependent transport properties of SGSs and HMMs hold promise 
for novel spintronic devices. Recent proposals include a reconfigurable magnetic tunnel 
diode and transistor concept based on these materials \cite{sasioglu2019proposal,maji2022demonstration}.

SGMs represent a recently proposed concept in spintronics, introduced by the present authors \cite{sasioglu2024spin}. 
This concept builds upon the idea of gapped metals, which are materials possessing a band gap 
slightly above or below the Fermi level \cite{Malyi2020,Ricci2020,Khan2023}. Gapped metals exhibit intrinsic 
p- or n-type conductivity unlike conventional semiconductors requiring extrinsic doping. Similarly, SGMs are 
predicted to display intrinsic p- or n-type behavior for each spin channel independently. Their properties 
would be similar to the dilute magnetic semiconductors eliminating the requirement for transition metal 
doping \cite{Sato2010,Lei2022,Tu2014,Kroth2006}. In Fig.\ref{fig1} (f-j) we present five distinct scenarios 
of the schematic DOS  for spin gapped metals. By combining p- or n-type gapped metallic behavior for the 
spin-up electronic band structure with various behaviors such as normal metallic, typical semiconducting, 
or n(p)-type gapped metallic behavior in the spin-down electronic band structure, a broader range of implications 
for device applications could be achieved as will be discussed in the following section.

As shown in Fig.\,\ref{fig1} SGSs can be qualitatively described by two band gap parameters: 
the internal band gap, denoted by $E_{\mathrm{g}}^{\mathrm{I}}$, and the external band gap, 
$E_{\mathrm{g}}^{\mathrm{E}}$. Type-I and type-III SGSs possess an external and internal band 
gap, respectively. In contrast, type-II SGSs exhibit both types of band gaps. Type-IV SGSs 
can be regarded as considered half-metallic magnets. As discussed later, these band gaps are 
crucial in determining the transfer ($I_D$-$V_G$) characteristics of SGS-based FETs. SGMs
require two additional parameters for the characterization of their electronic structures: 
$W_{\mathrm{m}}^{\mathrm{E}}$ and W$_{\mathrm{m}}^{\mathrm{I}}$. $W_{\mathrm{m}}^{\mathrm{E}}$ 
represents the energy difference between the Fermi level and the valence band maximum (or 
external band gap edge) for p-type SGMs, while $W_{\mathrm{m}}^{\mathrm{I}}$ represents the 
energy difference between the conduction band minimum (or internal band gap edge) and the 
Fermi level for n-type SGMs.

\section{Multifunctional  spintronic FETs}

\begin{figure}
\centering
\includegraphics[width=0.498\textwidth]{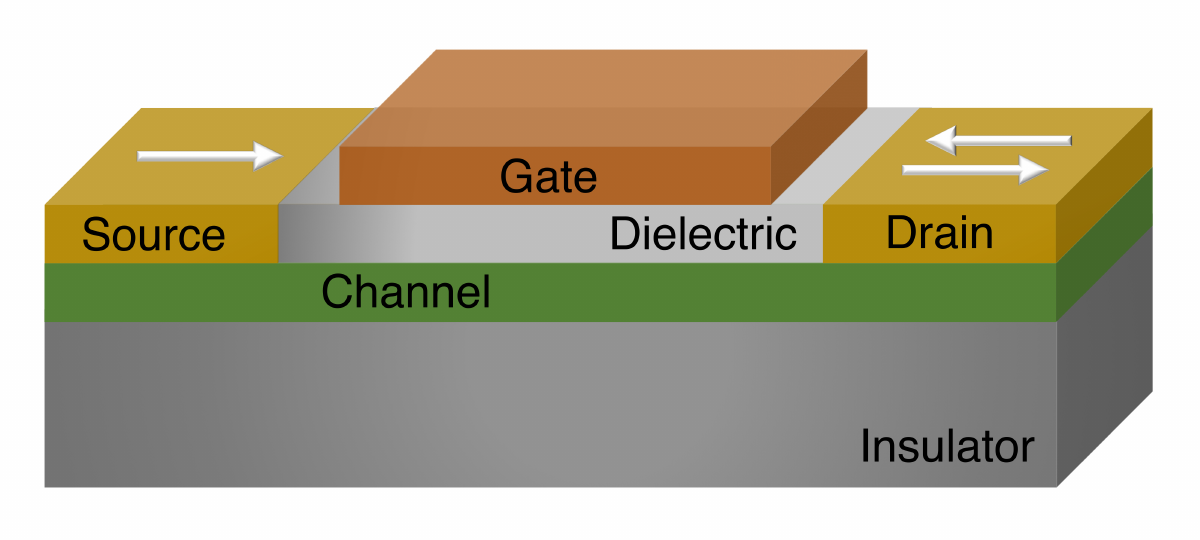}
\vspace{-0.6 cm}
\caption{Schematic representation of a two-dimensional field-effect transistor with spin gapless 
semiconductor and/or spin gapped metal source/drain electrodes. Arrows indicate magnetization 
direction.}
\label{fig2}
\end{figure}

In this section we introduces a novel device concept: multifunctional spintronic  FETs as illustrated 
in Fig.\,\ref{fig2}. The design leverages SGSs and/or SGMs for the source and drain electrodes, 
whose magnetization direction can be configured as either parallel or antiparallel. The channel, in 
contrast, is formed by an intrinsic semiconductor material. This unique combination creates a structure 
similar to a Schottky-barrier FET, where a potential barrier forms between the SGS (SGM) electrodes and 
the intrinsic channel. Electron injection in these devices occurs primarily through two mechanisms: i)
thermionic emission: Electrons with sufficient thermal energy overcome the Schottky barrier height and 
are injected into the channel. ii) tunneling: A small portion of electrons can tunnel through the barrier 
at lower energy levels. The interplay between these mechanisms and the relative alignment of the SGS (SGM) 
electrode magnetization influences the device's transfer characteristics ($I_D$-$V_G$). By exploring 
these $I_{\mathrm{D}}$-$V_{\mathrm{G}}$ characteristics we aim to understand the potential for manipulating 
spin currents and achieving functionalities beyond those of conventional FETs.

\begin{figure*}[t]
\centering
\includegraphics[width=0.999\textwidth]{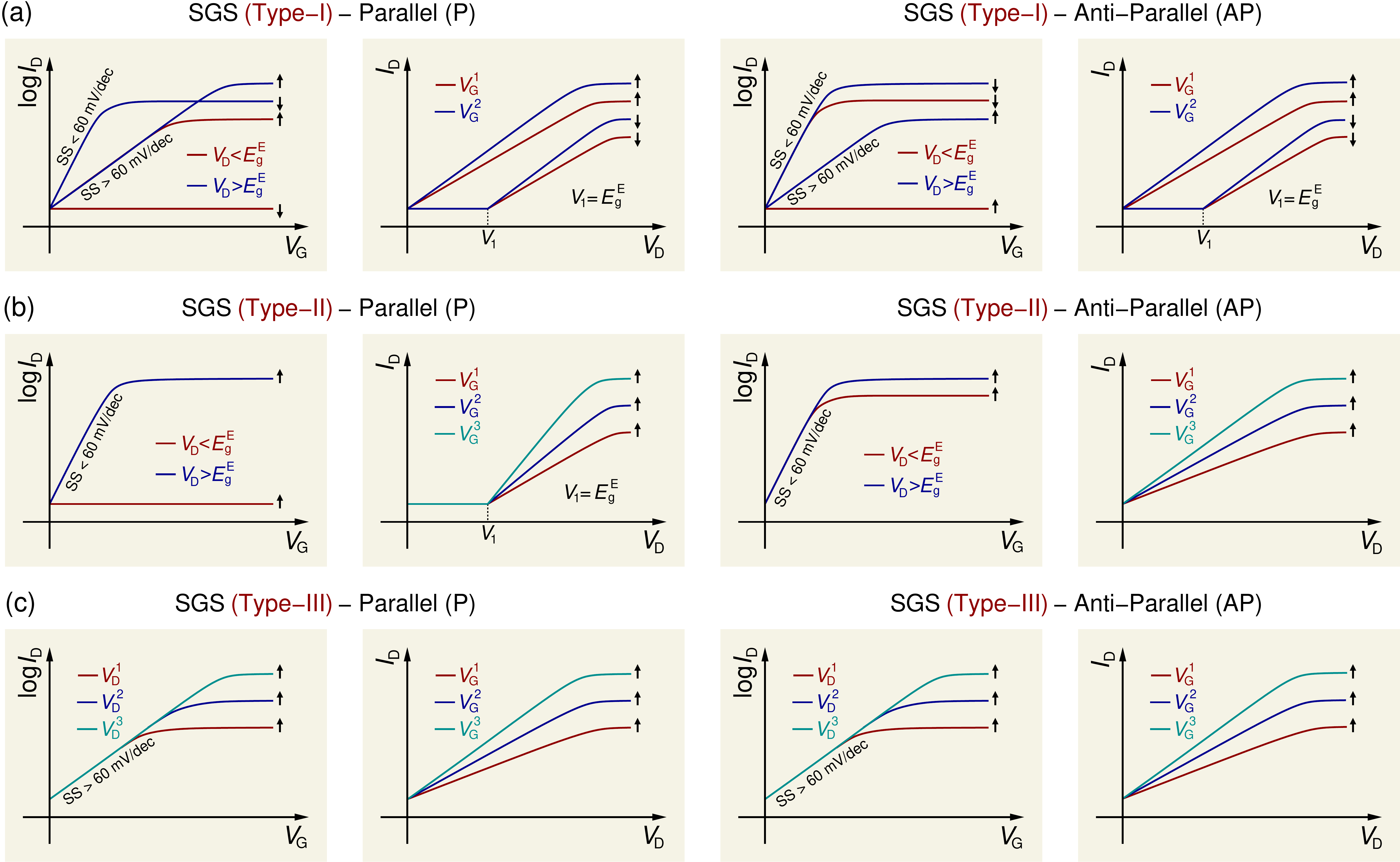}
\vspace{-0.2 cm}
\caption{(a) Schematic representation of the transfer ($I_{\mathrm{D}}$-$V_{\mathrm{G}}$ and output 
($I_{\mathrm{D}}$-$V_{\mathrm{D}}$) characteristics of a FET based on type-I SGS electrodes. 
The $I_{\mathrm{D}}$-$V_{\mathrm{G}}$ and $I_{\mathrm{D}}$-$V_{\mathrm{D}}$ curves are shown for two 
configurations of the source and drain electrode magnetization directions: parallel and antiparallel. 
The left-most panel shows the dependence of drain current ($I_{\mathrm{D}}$) on gate voltage ($V_{\mathrm{G}}$) 
for the majority (spin-up arrow) and minority (spin-down arrow)  spin channels at two different drain 
bias voltages $V_{\mathrm{D}}$. The adjacent panel illustrates the $I_D$ dependence on drain voltage 
($V_{\mathrm{D}}$) for both spin channels for two different gate voltage $V_{\mathrm{G}}$. The same 
configurations are repeated on the right for the anti-parallel magnetization case. (b) and (c) depict 
similar ($I_{\mathrm{D}}$-$V_{\mathrm{G}}$ and output ($I_{\mathrm{D}}$-$V_{\mathrm{D}}$) characteristics 
for FETs based on type-II and type-III SGS electrodes, respectively. SS denotes subthreshold slope.}
\label{fig3}
\end{figure*}

In Fig.\,\ref{fig3} we present the schematic the transfer($I_{\mathrm{D}}$-$V_{\mathrm{G}}$ and 
output ($I_{\mathrm{D}}$-$V_{\mathrm{D}}$) characteristics for FETs based solely on SGS electrodes. 
Here, we consider three types of SGSs (type-I, type-II, and type-III) and explore the influence of 
the relative magnetization orientation of the source and drain electrodes (parallel and antiparallel) 
on the drain current. Each SGS type is represented by four panels in Fig.\,\ref{fig3}, with two panels 
depicting the drain current ($I_{\mathrm{D}}$) as a function of gate voltage ($V_{\mathrm{G}}$) and 
the other two depicting $I_{\mathrm{D}}$ as a function of drain voltage ($V_{\mathrm{D}}$). These panels 
differentiate between the spin-up and spin-down current components for a comprehensive understanding.

Type-I SGSs possess states below the Fermi level in both spin channels, enabling current conduction 
in both.  The first panel of Fig.\,\ref{fig3} (a) presents the drain current ($I_{\mathrm{D}}$) versus 
gate voltage ($V_{\mathrm{G}}$) (transfer) characteristics for each spin channel with parallel source 
and drain magnetization. The spin-up channel exhibits conventional FET behavior with a sub-threshold 
swing (SS) exceeding 60 mV/dec. However, the spin-down channel displays a distinct characteristic.
For drain voltage ($V_{\mathrm{D}}$) lower than the external bandgap of the type-I SGS ($V_{\mathrm{E}} 
< E_{\mathrm{g}}^{\mathrm{E}}$), the drain current ($I_{\mathrm{D}}$) remains zero. This occurs 
because the drain electrode lacks unoccupied states to accommodate the incoming electrons from the 
source. Conversely, when $V_{\mathrm{D}}$ exceeds $E_{\mathrm{g}}^{\mathrm{E}}$, the spin-down 
$I_{\mathrm{D}}$ increases exponentially with an SS lower than 60 mV/dec before saturating. This 
steep-slope behavior originates from the unique spin-dependent band structure of the type-I SGS 
source electrode.  High-energy "hot" electrons in the spin-down channel of the source electrode are 
filtered out, leading to a reduced SS value in the sub-threshold region. In the on-state, the total 
$I_{\mathrm{D}}$ is the sum of spin-up and spin-down channel currents. Notably, in the sub-threshold 
region, the overall SS is expected to be lower than 60 mV/dec due to the filtering of high-energy 
electrons in the spin-down channel. The second panel of Fig.\,\ref{fig3}(a) shows the output   
($I_{\mathrm{D}}$-$V_{\mathrm{D}}$) characteristics for fixed gate voltages ($V_{\mathrm{G}}^1$ 
and $V_{\mathrm{G}}^2$) for the spin-up and spin-down channels. For the spin-up channel, $I_{\mathrm{D}}$ 
increases linearly with $V_{\mathrm{D}}$.  In contrast, the spin-down current remains zero until 
a critical drain-source bias voltage $V_1$, corresponding to the external bandgap ($V_1 = E_{\mathrm{g}}^{\mathrm{E}}$). 
These distinct characteristics highlight the interplay between spin-dependent transport and gate 
voltage control in type-I SGS FETs.  Similar transfer ($I_{\mathrm{D}}$-$V_{\mathrm{G}}$) and output 
($I_{\mathrm{D}}$-$V_{\mathrm{D}}$) characteristics can be observed for the anti-parallel magnetization 
configuration, as shown in the third and fourth panels of Figure \ref{fig3}(a).

\begin{figure*}[t]
\centering
\includegraphics[width=0.99\textwidth]{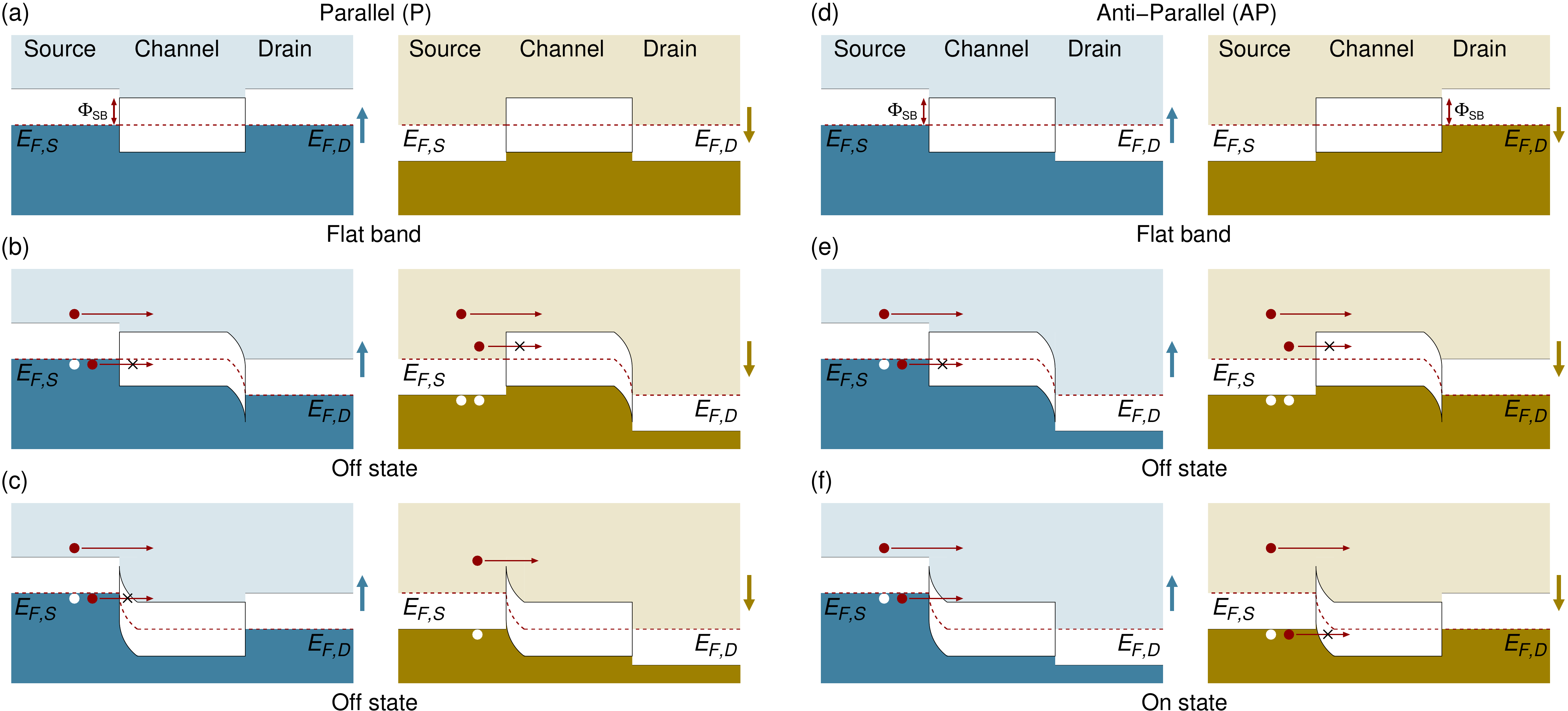}
\vspace{-0.01 cm}
\caption{Schematic representation of the spin-resolved energy-band diagram for the SGS/I/HMM MTJ
for parallel (P) orientation of the magnetization directions of the electrodes (a) for zero bias, (b) under
forward bias, and (c) under reverse bias. The electrons (holes) and the Fermi energy are denoted by
red (white) spheres and a dashed line, respectively, and the tunneling process is illustrated by partly
shaded red arrows. Panels (d)–(f) represent the same as (a)–(c) for the antiparallel (AP) orientation of
the magnetization directions of the electrodes.}
\label{fig4}
\end{figure*}

Unlike their type-I counterparts, type-II SGSs possess an internal gap in the opposite spin channel, 
as illustrated in Fig.\,\ref{fig1}. Consequently, only the spin-up channel can contribute to the drain 
current. The resulting transfer and output characteristics for type-II SGS FETs (Figure \ref{fig3}(b)) resemble 
those observed for the spin-down channel in type-I SGS devices. For transistor operation in the parallel 
magnetization configuration (source and drain magnetizations aligned), a drain bias voltage exceeding 
the external bandgap ($E_{\mathrm{g}}^{\mathrm{E}}$) is necessary. Conversely, the anti-parallel configuration 
(antiparallel magnetization configuration) does not require this condition (see Fig.\,\ref{fig4}). The 
external bandgap of the type-II SGSs materials considered in this work is greater than 0.5 eV. Therefore, 
for low-voltage operation, only the anti-parallel configuration is viable. In the parallel configuration, 
the transistor remains permanently off, leading to a 100\%  non-local giant magnetoresistance (GMR) effect, 
which will be discussed later.

In contrast to type-I and type-II SGSs, type-III SGSs exhibit a distinct band structure. They lack a 
bandgap above the Fermi level, leading to conventional transfer characteristics in FETs based on these 
materials. As shown in Fig,\,\ref{fig3}, the  SS value exceeds 60 mV/dec. However, a bandgap exists 
below the Fermi level in the spin-down channel of these materials, enabling 100\% spin-polarized drain 
current and a non-local GMR effect.

The non-local GMR effect can be defined as $\mathrm{GMR}=\left(I_{\mathrm{D}}^{\mathrm{P}} - 
I_{\mathrm{D}}^{\mathrm{AP}}\right)/\left(I_{\mathrm{D}}^{\mathrm{P}} + 
I_{\mathrm{D}}^{\mathrm{AP}}\right)$, where $I_{\mathrm{D}}^{\mathrm{FM}}$ 
$I_{\mathrm{D}}^{\mathrm{AP}}$) is the drain current in the parallel (anti-parallel) 
orientation of the magnetization of the source-drain electrodes. While theory predicts a maximum 
non-local GMR of 100\% at zero temperature for a FET with ideal SGS electrodes and a channel 
lacking spin-orbit coupling, real materials exhibit deviations from this ideal SGS behavior. 
For example, type-II SGS materials have overlapping valence and conduction bands, resulting in 
spin gapped metallic behavior that can reduce the GMR effect. Additionally, finite temperatures 
introduce thermal excitations (hot electrons) that further diminish the GMR. We will delve deeper 
into this temperature dependence using the energy-band diagram of a FET based on type-II SGS 
electrodes.

The energy-band diagrams in Fig.\,\ref{fig4} illustrate the operating principles of a 
FET with type-II SGS source and drain electrodes and an intrinsic semiconductor channel. These diagrams 
depict the changes in the energy bands for both spin channels under three distinct biasing conditions, 
considering both parallel and anti-parallel magnetization alignments of the source-drain electrodes.
In the absence of any applied bias (flat-band condition), the Fermi levels of the SGS electrodes align 
with the intrinsic level of the semiconductor, resulting in a flat energy band profile across the device 
for both spin channels. When a positive bias voltage is applied to the drain electrode relative to the 
source (off-state), the energy bands of the semiconductor bend downwards in the vicinity of the drain. 
In this off-state, for sufficiently long channel lengths, minimal current (tunneling leakage current) 
flows through the device for either spin channel. Finally, applying a positive bias to both the drain 
and the gate electrode (on-state) induces a downward shift in the energy bands of the channel. This allows 
for spin-up electrons in the source electrode to tunnel through the Schottky barrier and flow into the 
channel, eventually reaching the drain electrode, turning the FET "on". Note that  in Fig.~\ref{fig4}, 
the energy-band diagrams are drawn under the assumption that the drain bias voltage 
($V_{\mathrm{D}}$)  is smaller than the external bandgap ($E_{\mathrm{g}}^{\mathrm{E}}$) of the type-II 
SGS electrodes. This ensures the transistor is in the off-state for parallel electrode magnetization. 
In this configuration, the absence of unoccupied states in the drain electrode prevents carrier injection. 
Conversely, for the anti-parallel orientation shown in Fig.~\ref{fig4}(f), only spin-up electrons can 
be injected from the source to the drain, resulting in a spin-polarized current.

The steep slope behavior of the FET based on type-II SGSs (see $I_{\mathrm{D}}$-$V_{\mathrm{G}}$ transfer 
characteristics in Fig.\,\ref{fig3}) originates from the presence of a bandgap above the Fermi level in 
the spin-up channel. This energy gap filters out the thermally excited high-energy hot electrons giving 
rise to an electron injection within a very narrow energy window at the Fermi level from the source electrode 
into the drain. The filtering efficiency depends on the magnitude of the external bandgap $E_{\mathrm{g}}^{\mathrm{E}}$.  
A larger bandgap leads to a more stringent filtering effect, blocking higher-energy electrons. In materials that 
we consider in this work the value of the external gap $E_{\mathrm{g}}^{\mathrm{E}}$ ranges from 0.5 eV 
to 0.9 eV. Only a small number of very high-energy thermally excited electrons, residing at the tail of 
the Fermi-Dirac distribution can tunnel through the bandgap or thermally injected from either spin channel 
as illustrated in the energy-band diagrams of Fig.~\ref{fig4}. These "hot electrons" 
contribute to the leakage current in the off-state, diminish the non-local GMR effect, and increase the 
transistor's SS value.

\begin{figure*}[t]
\centering
\includegraphics[width=0.999\textwidth]{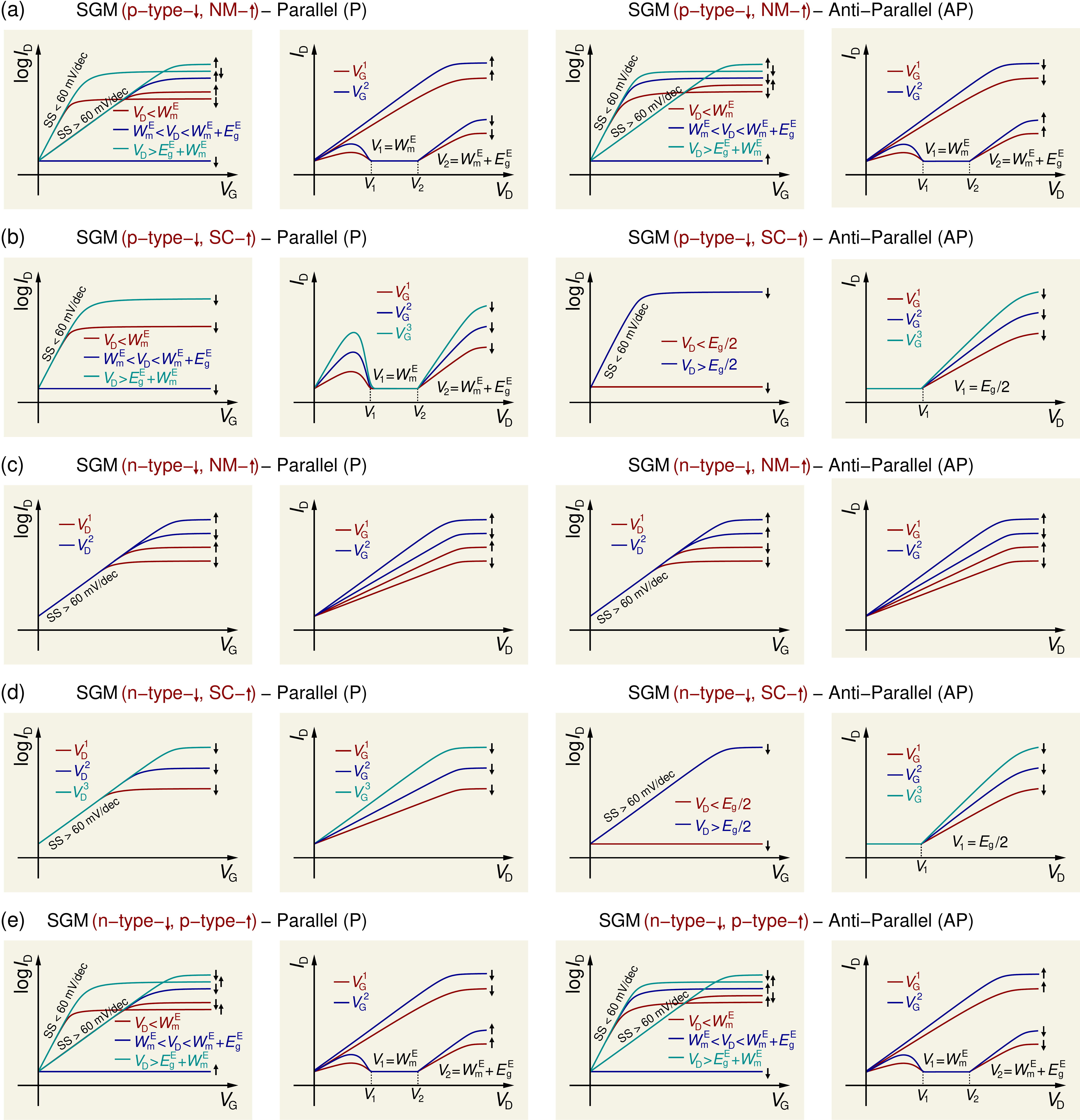}
\vspace{-0.2 cm}
\caption{ (a) Schematic representation of the transfer ($I_{\mathrm{D}}$-$V_{\mathrm{G}}$ and 
output ($I_{\mathrm{D}}$-$V_{\mathrm{D}}$) characteristics of a FET based on p-type SGM electrodes. 
The $I_{\mathrm{D}}$-$V_{\mathrm{G}}$ and $I_{\mathrm{D}}$-$V_{\mathrm{D}}$ curves depict two 
configurations of source and drain electrode magnetization directions: parallel and anti-parallel.  
The left-most panel shows the dependence of drain current $I_{\mathrm{D}}$ on gate voltage 
$V_{\mathrm{G}}$ for the majority (spin-up arrow) and minority (spin-down arrow)  spin channels 
at three different drain bias voltages $V_{\mathrm{D}}$. The adjacent panel illustrates the 
$I_{\mathrm{D}}$ dependence on drain voltage ($V_{\mathrm{D}}$) for both spin channels for two 
different gate voltage $V_{\mathrm{G}}$. The same configurations are repeated on the right for 
the anti-parallel magnetization case. (b) The transfer ($I_{\mathrm{D}}$-$V_{\mathrm{G}}$) and output 
($I_{\mathrm{D}}$-$V_{\mathrm{D}}$) characteristics of a similar FET with p-type SGM electrodes, 
but where only the spin-down channel contributes to the drain current. (c) and (d)  depict similar 
transfer and output characteristics for FETs based on n-type SGMs, following the same panel descriptions 
as provided for (a). (e) Transfer and output characteristics of a FET based on SGM electrodes with 
a p-type spin-up channel and n-type spin-down channel. SS denotes subthreshold slope.}
\label{fig5}
\end{figure*}

Both type-I and type-II SGS-based transistors exhibit low-voltage operation 
achieving SS values below 60 mV/dec, provided the drain voltage is smaller than the external 
bandgap of the SGS electrodes. However, a key difference exists between the two types. Type-I 
SGS-based FETs function regardless of the relative magnetization direction of the electrodes 
(as shown in Fig.~\ref{fig3}). Conversely, type-II SGS-based FETs only exhibit transistor behavior 
when the electrodes are in an anti-parallel orientation. This limitation in type-II devices is 
compensated by a significantly stronger non-local GMR effect. It is important to note that 
for type-II SGSs, the spin-gapless semiconducting properties lack inherent symmetry protection 
and can only emerge when a free parameter, such as pressure, is tuned to a specific value. 
Consequently, an ideal 2D free-standing type-II SGS material might lose its SGS properties 
when integrated with a dielectric substrate, as in field-effect transistors (FETs), or when 
it forms a heterojunction with other 2D materials. In these scenarios, either an overlap of 
spin-up valence and spin-down conduction bands or a shift of the Fermi level into the valence 
or conduction band occurs, leading to a new class of materials we call spin gapped metals (see 
Fig.~\ref{fig1}). We will discuss the $I$-$V$ characteristics of FETs based on these spin-gapped 
metals in the following.

Our recent paper comprehensively explored the concept of SGMs. While seven distinct types were 
identified, in this work we focus on five key cases (illustrated in Fig.\,\ref{fig1}). 
We specifically exclude materials in FET design exhibiting p-type or n-type spin gapped 
behavior in both spin channels. The first considered spin-gapped metal resembles a type-I 
SGS with a Fermi level shifted into the valence band [Fig.~\ref{fig1}(f)]. This configuration 
leads to metallic behavior in the spin-up channel, while the spin-down channel exhibits 
p-type characteristics. The transfer ($I_{\mathrm{D}}$-$V_{\mathrm{G}}$) characteristics of the 
corresponding FET presented in Fig.\,\ref{fig5} (a) resemble those of type-I SGS-based FETs, 
but with a crucial distinction in the number of distinct drain-source bias voltage regions. 
Unlike type-I SGS FETs (which have two), this device exhibits three. For very small drain biases 
($V_{\mathrm{D}} < W_{\mathrm{m}}$), both spin channels contribute to the current [Fig.\,\ref{fig5} (a)]. 
In the intermediate and high bias region ($W_{\mathrm{m}} < V_{\mathrm{D}} < W_{\mathrm{m}} + E_{\mathrm{g}}$ 
and $V_{\mathrm{D}} > W_{\mathrm{m}} +E_{\mathrm{g}}$), the transfer characteristics are similar 
to type-I SGS FETs. A key advantage of FETs based on p-type spin-gapped metals is the emergence of 
a NDR effect in the spin-down channel, as shown in the second panel of Figure 5(a). This behavior
contrasts sharply with the linear response observed in SGS-based FETs. For a sufficiently large 
fixed gate voltage, increasing the drain bias voltage ($V_D$) in Fig.\,\ref{fig5} (a) causes the 
spin-down channel current to initially rise to a peak value. However,  the current then decreases, 
eventually reaching zero at $V_{\mathrm{D}} = W_{\mathrm{m}}$ (denoted as $V_1$ here). The current 
remains zero until another critical voltage, $V_{\mathrm{D}} = W_{\mathrm{m}} + E_{\mathrm{g}}$ 
(denoted as $V_2$ here), is reached, after which it resumes a linear increase. In contrast, the spin-up 
channel current exhibits the typical linear behavior 
observed in type-I SGS FETs.

The next SGM we consider as an electrode material exhibits a p-type character, but unlike the 
previous case, one spin channel behaves as a semiconductor instead of a metal. In this FET, only 
the spin-down channel contributes to the drain current, resulting in transfer ($I_{\mathrm{D}}$-$V_{\mathrm{G}}$ 
and output ($I_{\mathrm{D}}$-$V_{\mathrm{D}}$) characteristics [Fig.~\ref{fig5}(b)] very similar 
to those in Fig.~\ref{fig5}(a) for parallel orientation of the electrode magnetization directions. 
However, for anti-parallel orientation, the semiconducting gap in the spin-up channel of the SGM 
electrode plays a crucial role in determining the FET's transfer characteristics. At low drain bias 
voltage ($V_{\mathrm{D}}< E_{\mathrm{g}}/2 $), this FET offers a key advantage: a very high non-local GMR effect.

In contrast to FETs based on p-type SGMs, those based on n-type SGMs exhibit normal transistor 
functionality similar to conventional FETs, i.e., with SS values exceeding 60 mV/dec. Fig.\,\ref{fig5}(c) 
and (d) present the corresponding transfer ($I_{\mathrm{D}}$-$V_{\mathrm{G}}$ and output 
($I_{\mathrm{D}}$-$V_{\mathrm{D}}$) characteristics of such FETs based on n-type SGM with spin-up 
channel metallic and spin-up channel semiconducting, respectively. In the latter case, only the 
spin-down channel contributes to the drain current, which is therefore spin-polarized. Similar to 
p-type FETs with one spin channel semiconducting SGM, these n-type FETs also exhibit a very high 
non-local GMR effect.

The final SGM we consider exhibits p-type behavior in the spin-up channel and n-type behavior 
in the spin-down channel. This p-n type SGM can be conceptualized as a type-II SGS where the 
conduction and valence bands overlap. Fig.\,\ref{fig5}(e) presents the transfer 
($I_{\mathrm{D}}$-$V_{\mathrm{G}}$ and output ($I_{\mathrm{D}}$-$V_{\mathrm{D}}$) characteristics 
of FETs based on these SGMs. As can be seen, the schematic $I_{\mathrm{D}}$-$V_{\mathrm{G}}$ and  
$I_{\mathrm{D}}$-$V_{\mathrm{D}}$ curves are similar to those of FETs based on p-type SGMs with a 
metallic spin-up channel.  Similar to the previous cases, three distinct drain bias voltage regions 
are observed. For very small drain biases ($V_{\mathrm{D}} < W_{\mathrm{m}}$), 
both spin channels contribute to the current [Fig.\,\ref{fig5}(e)]. In this regime, the transistor 
exhibits a gate-tunable NDR effect in the spin-up (spin-down) channel for parallel (anti-parallel) 
alignment of the electrodes' magnetization direction. Notably, the non-local GMR effect is 
observable across all drain bias voltages. It reaches a particularly high value within a specific 
voltage interval between $V_1$ and $V_2$. In this interval, the current in one spin channel 
diminishes due to the presence of a gap [as seen in the second and fourth panels of Fig.\,\ref{fig5}(e)].
These FETs based on SGM electrodes exhibit multifunctional characteristics, making them highly 
attractive for next-generation spintronic devices. Notably, they demonstrate sub-60 mV/dec 
switching, a hallmark of low-power electronics, alongside the non-local GMR effect for efficient 
spin manipulation and the gate-tunable NDR effect for potential applications in oscillators 
and high-frequency electronics.

Note that unlike conventional MOSFETs with bidirectional current flow, the proposed 
spintronic FETs exhibit an interesting dependence on the source and drain electrode 
materials. While the focus of this work lies on devices with identical SGS or SGM 
electrodes for both source and drain, the spin-dependent nature of these materials 
offers potential for exploring current directionality. Utilizing different SGS or 
SGM materials for the source and drain, as considered in our broader research, could 
potentially introduce a preference for current flow in one direction. This behavior 
aligns with our previously proposed reconfigurable magnetic tunnel diode \cite{sasioglu2019proposal,aull2022ab}. 
However, a detailed investigation of these mixed-material scenarios would require further study 
and might be the subject of future research. In the current work, the identical source 
and drain materials combined with the intrinsic channel create a more MOSFET-like behavior,
where the gate voltage primarily controls the overall conductance, enabling bidirectional 
current flow and on/off switching functionality.

Having established the potential of FETs based on SGS and SGM electrodes through analysis of 
their schematic transfer and output characteristics, we now delve into a more comprehensive 
exploration using computational methods. The following sections detail the chosen computational 
approach, the screening process for identifying promising 2D SGS and SGM materials, and finally, 
quantum transport calculations performed on a vertical FET employing a selected SGS material. This 
multifaceted approach allows us to move beyond schematic representations and validate the 
functionality of these novel devices through rigorous computational simulations.

\section{Computational Method \label{sec:Method}}

We employed density functional theory (DFT) implemented in the QuantumATK package \cite{QuantumATK,smidstrup2019an}  
to calculate the ground-state electronic structure of the materials. The GGA-PBE exchange-correlation 
functional \cite{perdew1996generalized} was used in conjunction with Pseudo-Dojo pseudopotentials 
\cite{QuantumATKPseudoDojo} and LCAO basis sets. A dense k-point grid of $24 \times 24 \times 1$ and a density 
mesh cutoff of 120 Ha were employed. To eliminate interactions between periodic images, a vacuum 
layer of 20\,{\AA} was added, and Neumann boundary conditions were applied. Convergence 
criteria for total energy and forces were at least 10$^{-4}$\,eV and 0.01\,eV/{\AA}, respectively.

Transport calculations were performed using a combination of DFT and the nonequilibrium Green 
function method (NEGF) implemented within QuantumATK.  A dense $\mathbf{k}$-point grid of $24 
\times 1 \times 172 $ was employed for self-consistent DFT-NEGF calculations. The transfer and output  
characteristics were obtained using the Landauer approach \cite{Landauer-Buettiker}, where the 
current is expressed as: $ I(V) = \frac{2e}{h}\sum_{\sigma}\int , T^{\sigma}(E,V)
\left[f_{L}(E,V)-f_{R}(E,V)\right] \mathrm{d}E $.  In this equation, $V$ represents the applied 
bias voltage, $T^\sigma (E,V)$ is the spin-dependent transmission coefficient for an electron with 
spin $\sigma$, and $f_L(E,V)$ and $f_R(E,V)$ are the Fermi-Dirac distribution functions for the 
left and right electrodes, respectively. The transmission coefficient, $T^\sigma (E,V)$, is calculated 
using a finer $\mathbf{k}$-point grid of $300 \times 1$.

\section{Results and Discussion \label{sec:ResandDis}}

The preceding section of this manuscript established the concept of multifunctional 
spintronic FETs through a conceptual foundation. We explored the transfer and output 
characteristics of these devices based on the schematic spin-resolved DOS and energy-band diagrams for 
SGSs and SGMs used as electrode materials. This section builds upon that foundation by presenting the 
results and discussion of our computational investigations. We will detail the screening 
process employed to identify promising 2D SGS and SGM candidates using DFT calculations. 
Following this, we will showcase the design and exploration of a specific vertical 
heterojunction FET example utilizing VS$_2$ (a type-II SGS) as the source and drain 
electrodes.

\subsection{Screening of 2D SGSs and SGMs }

\begin{table*}[!ht]
\caption{\label{table1}
Lattice constants ($a$, $b$), sublattice and total magnetic moments, magnetic anisotropy energy
(MAE), work function ($\Phi$), spin gap type per spin direction, internal and external energy 
gaps ($E_{\mathrm{g}}^{\mathrm{I/E}}$), formation energy ($E_\text{form}$), and convex hull 
distance energy ($\Delta E_\text{con}$) for the compounds under study. The $a$, $b$, MAE,  
$E_\text{form}$ and $\Delta E_\text{con}$ values are taken from the 2D Computational 
Materials Database~\cite{2dmd}.}
\begin{ruledtabular}
\begin{tabular}{lccllllllll}
Compound & $a$ & $b$ & m$_{\mathrm{TM}}$ & m$_\text{total}$  &  MAE & $\Phi$ & SGS & $E_{\mathrm{g}}^{\mathrm{I/E}}$ & $E_\text{form}$ &  $E_\text{con}$  \\
             & ({\AA}) & ({\AA}) & ($\mu_B$)   & ($\mu_B$) & (meV) & (eV) & Type & (eV) & (eV/at.) & (eV/at.)  \\ \hline
Ti$_4$Cl$_4$Te$_4$ & 7.07 & 7.07 & 1.13        & 4.00  & 1.47 (y) & 4.60 & Type-I  & \hspace{0.45 cm}/0.59 & -1.03 & 0.096  \\
V$_3$MoSe$_8$      & 6.67 & 5.77 & 1.04 (1.12) & 3.00  & 2.42 (y) & 5.33 & Type-I  & \hspace{0.45 cm}/0.72 & -0.68 & 0.026  \\
VSi$_2$N$_4$       & 2.88 & 2.88 & 1.17        & 1.00  & 0.10 (.) & 5.60 & Type-II &             1.73/0.58 & -0.95 &   \\
VS$_2$             & 3.18 & 3.18 & 1.08        & 1.00  & 0.21 (y) & 5.76 & Type-II &             0.78/0.75 & -0.88 & 0.000  \\
VSSe               & 3.26 & 3.26 & 1.18        & 1.00  & 0.40 (y) & 5.29 & Type-II &             0.56/0.50 & -0.79 & 0.010  \\
ScI$_2$            & 3.98 & 3.98 & 0.92        & 1.00  & 0.51 (y) & 3.47 & Type-II &             2.47/0.82 & -1.14 & 0.008  \\

\end{tabular}
\end{ruledtabular}
\end{table*}

Three-dimensional (3D) SGSs have been extensively studied both experimentally and theoretically
\cite{ouardi2013realization,aull2019ab,ozdougan2013slater,galanakis2016spin,xu2013new,gao2019high}. 
However, for next-generation low-power and miniaturized electronics, 2D materials offer 
significant advantages due to their unique properties and potential for device scaling 
\cite{geim2007rise,chhowalla2016two,allain2015electrical,wang2012electronics}. A crucial 
advantage of 2D materials in this context is the ability to tune the Schottky barrier height 
at the interface between the channel material and the electrodes using a gate voltage 
\cite{yang2012graphene,yu2013vertically,moriya2014large,qiu2015electrically,dankert2017electrical,lagasse2019gate}. 
This gate control over the barrier allows for more flexible device operation compared to 3D 
counterparts. Identifying suitable 2D SGSs and SGMs is crucial for realizing the multifunctional 
spintronic FETs proposed in this work. Ideal candidates should possess specific properties to achieve 
the desired functionalities. We conducted a search in the computational 2D materials database (C2DB) 
\cite{haastrup2018computational,gjerding2021recent} and identified 6 SGSs and 14 SGMs as outlined 
in Table\,\ref{table1} and Table\,\ref{table2}, respectively. Our initial criterion for selecting 
materials for the study was their formation energy, $E_{\text{form}}$, which we aimed to be negative. 
However, negative $E_{\mathrm{form}}$ alone does not guarantee stability. The convex hull distance 
$E_{\mathrm{con}}$, which represents the energy difference between the studied structure and the most 
stable phase or a mixture of phases, is also crucial. Typically, values less than 
0.1 eV/atom is desired to facilitate the growth of a material. All materials selected from 
C2DB for our study exhibits $E_{\text{con}}$ values less than the cutoff of 0.1 eV/atom.

While negative formation energy and low convex hull distance are essential for material 
stability and growth, another important consideration for SGS and SGMs is their Curie 
temperature $T_{\mathrm{c}}$. Ideally, SGS and/or SGM electrodes should exhibit ferromagnetism 
at or above room temperature for practical device applications. However, the primary focus 
of this work is to establish the proof-of-concept for our proposed multifunctional FETs. 
Therefore, we include materials like VS$_2$ with a theoretically estimated $T_{\mathrm{c}}$
below room temperature (around 120 K) to demonstrate the feasibility of the device concept \cite{fuh2016newtype}. 
On the other hand, materials such as VSi$_2$N$_4$ (a type-II SGS) have been reported to possess 
a $T_{\mathrm{c}}$ exceeding room temperature \cite{akanda2021magnetic}, which aligns with our 
preliminary calculations. Additionally, experimental observations of above-room-temperature 
ferromagnetism exist in 2H-VSe$_2$, a material with a band structure similar to type-II SGSs 
but with a small band gap\cite{wang2021ferromagnetism,jiang2020ferroelectric}. When VSe$_2$ forms 
a heterostructure with other 2D materials like MoS$_2$ or WS$_2$, charge transfer can potentially 
transform it into an SGM, likely retaining a $T_{\mathrm{c}}$ above room temperature.

The electronic and magnetic properties of the identified SGS materials in Table\,\ref{table1} 
are particularly attractive for FET applications. Of key importance are the internal and, 
especially, the external band gaps ($E_{\mathrm{g}}^{\mathrm{I/E}}$). A large external 
band gap is crucial for filtering out high-energy hot electrons in FETs, enabling sub-60 
mV/dec SS values as discussed in the preceding section. As shown in Table\,\ref{table1}, 
all considered compounds possess calculated external band gaps exceeding 0.5 eV.  Furthermore, 
the V-based compounds exhibit lattice parameters compatible with existing 2D semiconductors, 
such as MoS$_2$, MoSSe, MoSi$_2$N$_4$, and others \cite{hong2020chemical,mortazavi2021exceptional}. 
This compatibility facilitates seamless integration of these V-based SGS materials with established 
2D semiconductors within FET devices. It's important to note that while type-IV SGS behavior 
exists in some 2D materials \cite{sun2018prediction}, these materials are excluded from 
our analysis due to their lack of relevance for steep-slope FET applications.

\begin{table*}[!ht]
\caption{\label{table2}
Lattice constants $a$, $b$, sublattice and total magnetic moments, magnetic anisotropy energy
(MAE), work function $\Phi$, spin gap type per spin direction, the distance of the Fermi level 
from the edge of the band which it crosses $W_{\mathrm{m}}^{\mathrm{I,E}(\uparrow/\downarrow)}$ 
(see text for more details), energy gap per spin direction $E_{\mathrm{g}}^{\mathrm{I,E}(\uparrow/\downarrow)}$, 
formation energy ($E_\text{form}$), and convex hull distance energy ($\Delta E_\text{con}$) 
for the compounds under study. The $a$, $b$, MAE, $E_\text{form}$ and $\Delta E_\text{con}$ 
values are taken from the 2D Computational Materials Database~\cite{haastrup2018computational,gjerding2021recent}.} 
\begin{ruledtabular}
\begin{tabular}{lccllrllllll}
Compound & $a$ & $b$  & m$_{\mathrm{TM}}$ & m$_\text{total}$ & MAE & $\Phi$ & Spin gap type & $W_{\mathrm{m}}^{\mathrm{I,E}(\uparrow/\downarrow)}$ & $E_{\mathrm{g}}^{\mathrm{I,E}(\uparrow/\downarrow)}$ & $E_\text{form}$ &  $E_\text{con}$  \\
 & ({\AA}) & ({\AA}) & ($\mu_B$) & ($\mu_B$) & (meV)  &  (eV) & & (eV) & (eV) & (eV/at.) & (eV/at.)  \\ \hline
V$_3$MoS$_8$      & 6.37 & 5.51 & 0.49 (0.57)  & 1.46 & -0.10 ($z$) & 5.80 & p-type-$\uparrow$/p-type-$\downarrow$  & 0.34/0.75 & 0.85/0.94   & -0.87  & 0.020   \\
VOsO$_2$Br$_4$    & 5.24 & 5.24 & 1.49 (0.47)  & 1.99 &  3.70 ($y$) & 5.28 & p-type-$\uparrow$/p-type-$\downarrow$  & 0.21/0.47 & 0.59/0.88   & -0.96  & 0.036   \\
Cr$_2$Br$_2$Te$_2$& 5.46 & 3.76 & 3.54         & 6.10 & -2.95 ($z$) & 5.59 & \hspace{0.3 cm} NM-$\uparrow$/p-type-$\downarrow$  & \hspace{0.6 cm}/0.38 & \hspace{0.6 cm}/0.85 & -0.51 & 0.000  \\   
Cr$_2$I$_2$Te$_2$ & 5.43 & 3.93 & 3.55         & 6.05 & -3.13 ($z$) & 5.03 & \hspace{0.3 cm} NM-$\uparrow$/p-type-$\downarrow$  & \hspace{0.6 cm}/0.16 & \hspace{0.6 cm}/0.94 & -0.35 & 0.000  \\  
Cr$_3$Cl$_2$O$_4$ & 5.42 & 5.42 & 2.83 (2.99)  & 8.00 & -0.46 ($z$) & 5.84 &  p-type-$\uparrow$/SC-$\downarrow$                 &  0.15/               &   0.89/ & -1.61 & 0.090  \\  
PdCr$_2$Se$_4$    & 3.69 & 3.69 & 3.27         & 5.43 &  0.29 ($x$) & 5.44 & p-type-$\uparrow$/p-type-$\downarrow$  & 0.61/0.51 & 0.37/0.66            & -0.42   & 0.080   \\
VAl$_2$Se$_4$     & 3.90 & 3.90 & 2.86         & 2.99 &  0.25 ($y$) & 3.45 & \hspace{0.4 cm}  SC-$\uparrow$/n-type-$\downarrow$     & \hspace{0.6 cm}/0.11 & \hspace{0.6 cm}/1.39 & -0.76 & 0.040  \\
VIn$_2$S$_4$      & 3.79 & 3.90 & 2.92         & 2.96 &  0.14 ($x$) & 4.08 & \hspace{0.35 cm}  SM-$\uparrow$/n-type-$\downarrow$     & \hspace{0.6 cm}/0.24 & \hspace{0.6 cm}/1.52 & -0.65 & 0.070  \\   
Cr$_3$S$_4$       & 3.44 & 3.44 & 3.06 (-3.22) & 2.51 &  0.03 ($x$) & 5.20 & \hspace{-0.06 cm} n-type-$\uparrow$/n-type-$\downarrow$ & 0.37/0.34  & 0.29/0.43   & -0.66 & 0.050   \\
Cr$_4$F$_2$N$_3$  & 3.01 & 3.01 & 3.00 (-2.82) & 0.42 & -0.08 ($z$) & 5.11 & \hspace{0.30 cm} NM-$\uparrow$/n-type-$\downarrow$     & \hspace{0.6 cm}/0.18 & \hspace{0.6 cm}/0.74 & -1.09 & 0.030  \\
CrGa$_2$S$_4$     & 3.72 & 3.72 & 3.79         & 3.60 & -0.05 ($z$) & 3.71 & \hspace{-0.06 cm} n-type-$\uparrow$/n-type-$\downarrow$ & 0.27/0.48  & 1.12/1.68   & -0.64 & 0.050   \\
CrGa$_2$Se$_4$    & 3.92 & 3.92 & 3.95         & 3.80 & -0.19 ($z$) & 3.93 & \hspace{-0.06 cm} n-type-$\uparrow$/n-type-$\downarrow$ & 0.50/0.65  & 0.71/0.92   & -0.55 & 0.060   \\
CoGa$_2$S$_4$     & 3.61 & 3.61 & 0.91         & 1.00 &  0.07 ($x$) & 4.19 & \hspace{-0.06 cm}  n-type-$\uparrow$/SC-$\downarrow$     &  0.35/ &     0.56/ & -0.57 & 0.050  \\  
Mn$_2$Al$_2$S$_5$ & 3.68 & 3.68 & 4.61         & 9.75 & -0.11 ($z$) & 3.62 & \hspace{-0.06 cm} p-type-$\uparrow$/n-type-$\downarrow$ & 0.13/0.18  & 0.41/1.42   & -0.79 & 0.060   \\

\end{tabular}
\end{ruledtabular}
\end{table*}

Similar to the SGS materials discussed previously, the electronic properties of the SGMs presented 
in Table\,\ref{table2} are crucial for FET applications.  As with SGS, a large external band gap 
exceeding 0.5 eV is desired for efficient hot electron filtering and achieving sub-60 mV/dec SS values.  
All materials listed in Table\,\ref{table2} fulfill this criterion.  However, for SGMs, an additional 
parameter,  $W_{\mathrm{m}}^{\mathrm{I,E}(\uparrow/\downarrow)}$, comes into play. This parameter 
represents the energy difference between the valence (conduction) band edge and the Fermi level for 
p-type (n-type) SGMs, but crucially, it's specific to each spin direction (represented by the arrows). 
Table\,\ref{table2} reveals that some SGMs exhibit p-type or n-type character for both spin channels, 
while others exhibit mixed behavior with one spin channel being semiconducting or metallic and the other 
exhibiting p-type or n-type character. Only one 2D material in Table\,\ref{table2} presents the unique 
characteristic of p-type behavior for the spin-up channel and n-type behavior for the spin-down channel. 
For the p-type SGMs, the calculated $W_{\mathrm{m}}^{\mathrm{E}(\uparrow/\downarrow)}$ values range from 
0.16 eV to 0.75 eV, with similar values observed for n-type SGMs. As discussed earlier for p-type SGM-based 
FETs, the $W_{\mathrm{m}}^{\mathrm{E}(\uparrow/\downarrow)}$ parameter significantly impacts the NDR 
characteristics and the valley bias voltage in the $I_{\mathrm{D}}$-$V_{\mathrm{D}}$ curves.  While a larger 
$W_{\mathrm{m}}^{\mathrm{I,E}(\uparrow/\downarrow)}$ leads to a higher valley bias voltage, it also comes 
at the cost of higher SS values, as reported for gapped metals (or cold metals) in FETs \cite{yin2022computational}.  
Therefore, achieving an optimal balance between $W_{\mathrm{m}}^{\mathrm{E}(\uparrow/\downarrow)}$ and 
the external band gap, $E_{\mathrm{g}}^{\mathrm{E}(\uparrow/\downarrow)}$, is crucial for tailoring the 
transfer and output characteristics of FETs based on p-type SGMs.

The computational screening process successfully identified a range of promising 2D SGS and 
SGM candidates for multifunctional spintronic FETs. These materials exhibit a compelling 
combination of electronic and magnetic properties, including large external band gaps for 
efficient hot electron filtering.  The V-based SGS materials stand out due to their lattice 
parameters, which seamlessly integrate with existing 2D semiconductors, paving the way for 
straightforward device fabrication.  The next section takes a deeper dive by exploring a 
specific example: a FET design that utilizes VS$_2$ (a type-II SGS) as the source and drain 
electrodes. This investigation will further showcase the potential of these materials in 
realizing the proposed multifunctional FET concept.

\subsection{Vertical VS$_2$/Ga$_2$O$_2$ heterojunction FET}

\begin{figure*}[!ht]
\centering
\includegraphics[width=0.95\textwidth]{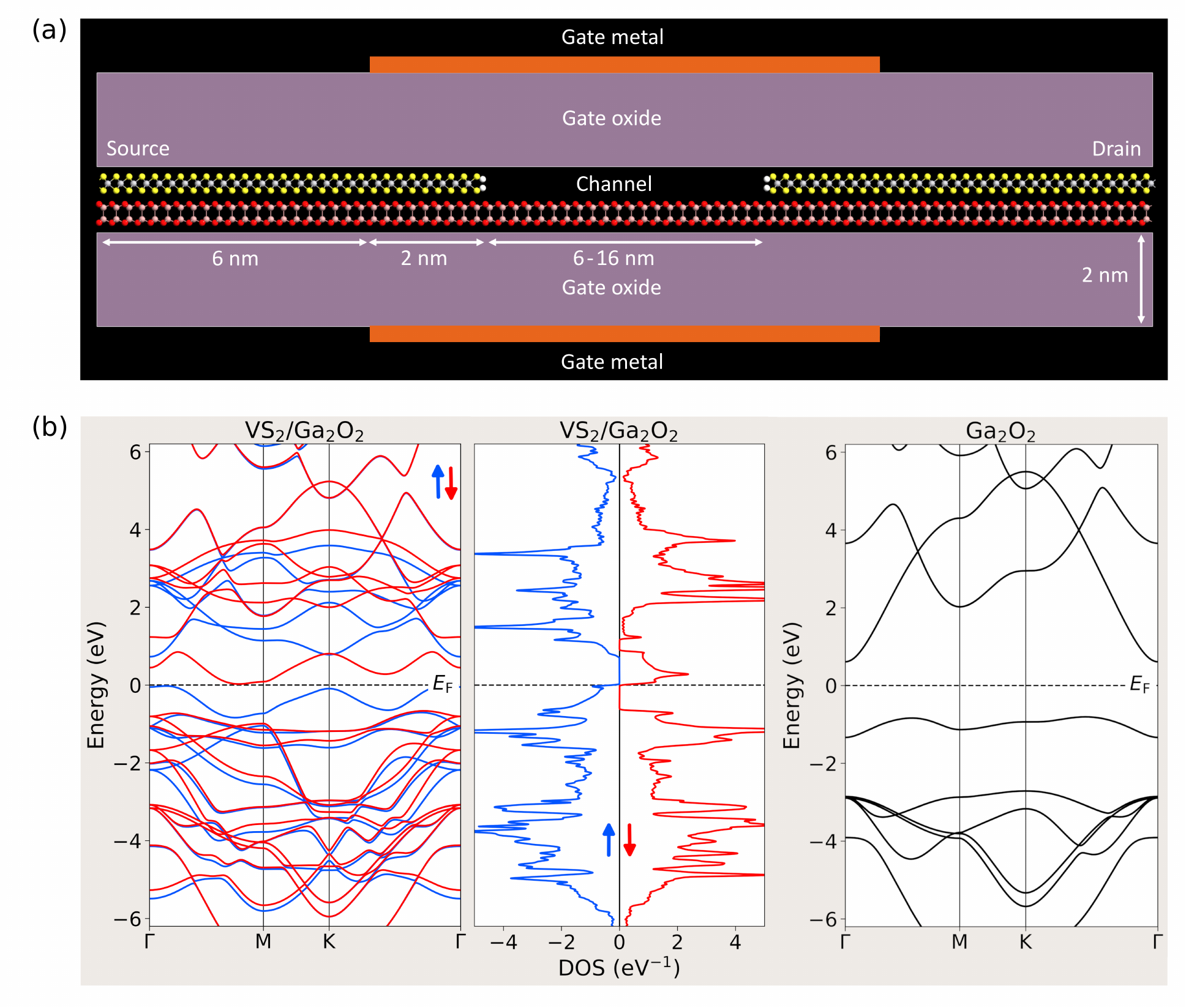}
\vspace{-0.2 cm}
\caption{(a) Schematic of the vertical VS$_2$/Ga$_2$O$_2$ heterojunction FET with key dimensions labeled. 
Channel length varies from 6 to 16 nm, while source/drain, gate overlap lengths, and gate oxide thickness 
are fixed at 6 nm, 2 nm, and 2 nm, respectively. (b) Calculated spin-resolved band structure along the high-symmetry 
directions and corresponding density of states (DOS) for the VS$_2$/Ga$_2$O$_2$ source/drain  (left panels) 
and the band structure of Ga$_2$O$_2$ channel material (right panel). The dashed black line indicates the Fermi 
level set to zero energy.}
\label{fig6}
\end{figure*}

Following the identification of promising 2D material candidates, this subsection explores 
a specific vertical heterojunction FET design utilizing VS$_2$ (a type-II SGS) as the 
source/drain electrodes and Ga$_2$O$_2$ as the channel material.  This configuration 
leverages VS$_2$'s properties to achieve the functionalities outlined earlier. Ga$_2$O$_2$ 
is chosen for the channel due to its small electron effective mass (0.33\,$m_0$), lattice 
matching, and comparable work function to VS$_2$. We first discuss the device geometry and 
then the electronic structure of source/drain and channel materials. The chosen device geometry 
adopts a vertical configuration with dual gates, as illustrated in Fig.\,\ref{fig6}. The source 
and drain electrodes (fixed at 6 nm) are composed of a VS$_2$/Ga$_2$O$_2$ heterostructure. The 
channel length varies from 6 to 16 nm, and the gate electrode overlaps the source and drain by 
2 nm with a 2 nm thick gate oxide (dielectric constant of 25). It's worth noting that the channel 
material, Ga$_2$O$_2$ (lattice constant: 3.13 \AA), has a slightly smaller lattice constant compared 
to VS$_2$ (3.18 \AA). To achieve a lattice-matched interface, the Ga$_2$O$_2$ experiences a slight 
in-plane strain (stretching) during the heterostructure formation.

\begin{figure*}[t]
\centering
\includegraphics[width=0.995\textwidth]{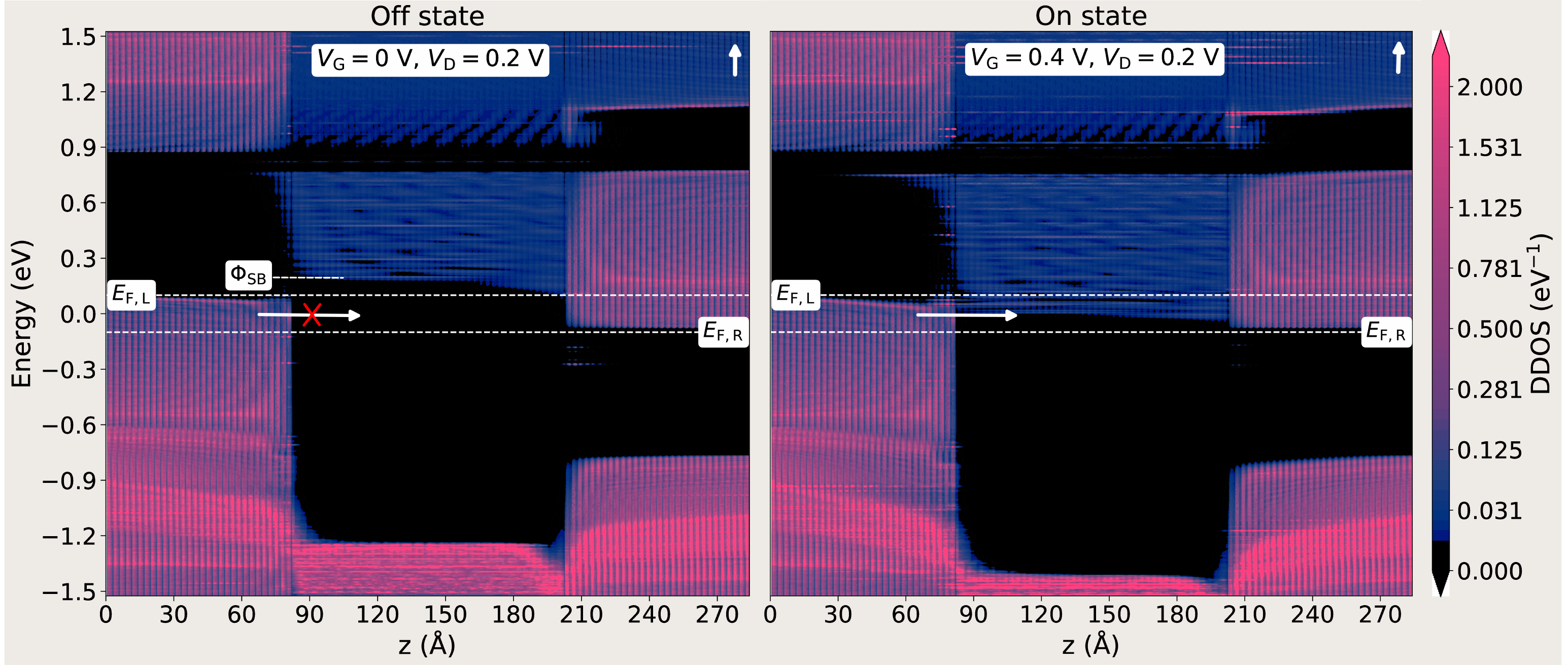}
\vspace{-0.4 cm}
\caption{Projected device density of states (DDOS) for majority spin (spin-up) electrons in a 12 nm channel 
length vertical VS$_2$/Ga$_2$O$_2$ heterostructure FET (see Fig.\,\ref{fig6} (a) for device structure). 
Left panel: off-state. Right panel: on-state. Upper and lower white dashed lines indicate the Fermi levels 
of the source and drain electrodes, respectively. $\Phi_{\mathrm{SB}}$ denotes the Schottky barrier.}
\label{fig7}
\end{figure*}

The variation in channel length, ranging from 6 to 16 nm, allows us to investigate the impact of 
channel length on subthreshold swing and leakage current. Due to the coherent transport mechanism, 
on-current is expected to be less dependent on channel length within this range. Shorter channels, 
however, can suffer from increased off-currents due to enhanced leakage currents through the channel. 
By exploring this range of channel lengths, we aim to identify an optimal balance between achieving a 
low SS value (indicating sharp switching) and minimizing leakage current for efficient device 
operation.

We opted for a VS$_2$/Ga$_2$O$_2$ heterostructure for the source and drain electrodes instead of a 
pure VS$_2$ monolayer. This choice maintains the type-II SGS character crucial for device applications, 
as evident from the spin-resolved band structure of the heterostructure in Fig.\,\ref{fig6}(b) and 
projected bands presented in the supplemental material \cite{SM}. Additionally, it simplifies the device design, 
reducing computational costs during simulations. As shown in Fig.\,\ref{fig6}(b), the vertical VS$_2$/Ga$_2$O$_2$ 
heterostructure exhibits ideal type-II SGS behavior, with occupied spin-up and unoccupied spin-down
bands aligned at the Fermi level. The channel material, Ga$_2$O$_2$, possesses a suitable bandgap 
of 1.56 eV and a favorable electron effective mass of  0.33\,$m_0$. However, the hole effective mass 
is significantly heavier at 3.13\,$m_0$. The calculated work function for the Ga$_2$O$_2$ channel 
($\Phi=6.36$\,eV) is higher compared to the source/drain electrodes ($\Phi=5.81$\,eV). This difference 
creates a Schottky barrier of 0.17\,eV at both the source-channel and channel-drain interfaces.

In supplemental material (Fig.\,S2), we present the spin-resolved local device density of states 
(DDOS) for a vertical VS$_2$/Ga$_2$O$_2$ heterostructure FET under anti-parallel electrode magnetization 
and flat-band conditions (zero gates and drain bias). The 12 nm channel DDOS clearly illustrates Schottky 
barriers at both the source/channel and channel/drain interfaces. Fig.\,\ref{fig7} depicts the spin-up 
channel DDOS for off-state ($V_{\mathrm{G}}=0$\,V, $V_{\mathrm{D}}=0.2$\,V) and on-state ($V_{\mathrm{G}}=0.4$\,V,
$V_{\mathrm{D}}=0.2$\,V) conditions, as the spin-up channel is the only current carrier in type-II SGS
materials. For completeness, the DDOS for the spin-down channel is presented in the supplemental material \cite{SM}. 
As seen in Fig.\,\ref{fig7}, off-state DDOS indicates a drain bias-induced barrier reduction near the 
channel-drain interface. In contrast, the on-state exhibits a gate-voltage-tuned Schottky barrier, decreasing 
from 0.17\,eV to 0\,eV. This reduction facilitates efficient electron injection from the source into the channel, 
contributing to a substantial drain current. It is worth noting that experimental observations of similar 
gate-voltage-induced Schottky barrier modulation have been reported in 2D material-based FETs 
\cite{yang2012graphene,yu2013vertically,moriya2014large,qiu2015electrically,dankert2017electrical,lagasse2019gate}.

\begin{figure*}[ht]
\centering
\includegraphics[width=0.995\textwidth]{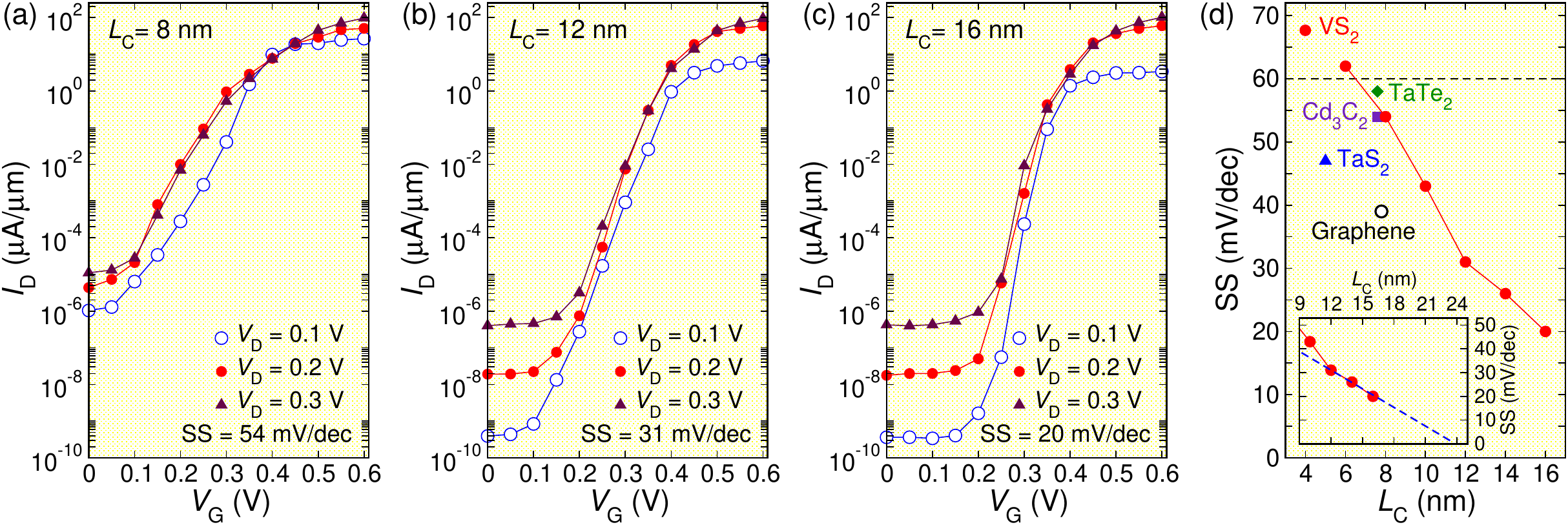}
\vspace{-0.6 cm}
\caption{The transfer characteristics ($I_{\mathrm{D}}$-$V_{\mathrm{G}}$) of the vertical VS$_2$/Ga$_2$O$_2$ heterostructure FET  
for three different source-drain bias voltages. (a-c) $I_{\mathrm{D}}$-$V_{\mathrm{G}}$ curves for different channel lengths 
($L_{\mathrm{C}}$) of 8, 12, and 16 nm, respectively. (d) Subthreshold slope (SS) as a function of channel length. The dashed 
line represents the thermionic limit. The SS values for cold-source FETs like TaS$_2$ are taken 
from Ref.\,\cite{wang2023cold}. Inset: Extrapolation of SS to longer channel lengths. }
\label{fig8}
\end{figure*}

The transfer characteristics ($I_{\mathrm{D}}$-$V_{\mathrm{G}}$) of the vertical VS$_2$/Ga$_2$O$_2$ 
heterostructure FET is calculated under antiparallel electrode magnetization configuration for three 
different source-drain bias voltages (0.1\,V, 0.2\,V, and 0.3\,V). As discussed in the preceding 
device concept section, this configuration enables low-voltage transistor operation in contrast to the 
parallel configuration. Channel length was varied from 6 nm to 16 nm in 2 nm increments.  Fig.\,\ref{fig8}  
presents representative $I_{\mathrm{D}}$-$V_{\mathrm{G}}$ curves  for channel lengths of 8 nm, 
12 nm, and 16 nm. The transistor exhibits a sharp transition from off-state to on-state within a narrow 
gate voltage range of 0.2-0.4\,V. For gate voltages exceeding 0.5\,V, the on-state current saturates at 
100 $\mu$A/$\mu$m. While saturated currents for drain voltages of 0.2\,V and 0.3\,V are comparable, 
off-state currents differ by over an order of magnitude, attributable to drain-induced barrier lowering. 
Fig.\,\ref{fig8}(d) illustrates the SS values as a function of channel length ($L_{\mathrm{C}}$), with SS values 
of cold-source 2D material FETs are included for comparison. The obtained SS value for a 6 nm channel 
slightly surpasses the room temperature thermionic limit of 60 mV/dec. However, SS decreases almost 
linearly with the increasing channel length, reaching 20 mV/dec at 16 nm. The inset of Fig.\,\ref{fig8}(d) 
shows SS extrapolation to longer channel lengths, indicating the potential for ideal transistor behavior (SS=0 
mV/dec) at around 24 nm. It is worth noting that threshold switch FETs based on 2D materials have reported 
an exceptionally low SS of 0.33 mV/dec \cite{TS}.

The on-state current density of the SGS FET is a key performance metric. For all investigated channel lengths, 
this value consistently reaches approximately 100 $\mu$A/$\mu$m when applying drain-source bias voltages of 0.2 
and 0.3\,V. Notably, this value drops by an order of magnitude at 0.1 V, likely due to wave-function mismatching 
between the source and drain electrodes. This observed current density significantly surpasses the much lower 
on-state values reported for tunnel FETs \cite{wang2023cold}, highlighting the potential advantages of our SGS FET 
architecture in terms of driving current capability. Furthermore, the device exhibits an on/off ratio exceeding 10$^8$ for 
channel lengths greater than 10 nm, demonstrating its excellent switching behavior. These combined characteristics 
position the SGS FET as a promising candidate for high-performance, low-power electronic applications, particularly 
where high current drive and steep subthreshold slope are essential.

SGSs can be regarded as the theoretical ultimate limit of cold-metals for steep-slope FETs \cite{zhang2024high,logoteta2020cold,wang2021intercalated}. 
As previously discussed, cold-source and cold-metal FETs hold promise for surpassing the conventional thermionic 
limit of 60 mV/dec for SS value at room temperature. However, reported SS values for cold-metal FETs remain 
relatively high \cite{wang2023cold,liu2020switching}. This limitation arises from the electronic band structure 
of the cold-metal electrodes. The energy gap edge ($W_{\mathrm{m}}^{\mathrm{E}}$) above the Fermi level [see 
Fig.\,\ref{fig1}(f) and (g)] and the external band gap ($E_{\mathrm{g}}^{\mathrm{E}}$) play a crucial role in 
filtering high-energy electrons. In cold metals with large $W_{\mathrm{m}}^{\mathrm{E}}$ values, only electrons 
in the deep subthreshold regime are effectively filtered, not those in the crucial subthreshold region. The 
cold-metal NbTe$_2$-based FET exemplifies this behavior \cite{liu2020switching}. Its intrinsic $W_{\mathrm{m}}^{\mathrm{E}}$ 
of 0.5 eV can be reduced to 0.27 eV by applying a 9\% strain, leading to an improved SS of 23 mV/dec in a 
NbTe$_2$-based FET. Type-I and type-II SGSs offer the potential to overcome this limitation by possessing an 
energy gap edge ($W_{\mathrm{m}}^{\mathrm{E}}$ ) approaching zero. Theoretically, this would enable ideal switching 
behavior with an SS of 0 mV/dec in an SGS-based FET under coherent transport conditions. Nevertheless, inelastic 
scattering processes, such as electron-phonon interactions, can still degrade the SS value in cold-metal FETs, as 
demonstrated theoretically \cite{duflou2021electron,afzalian2021ab}.

\begin{figure}[t]
\centering
\includegraphics[width=0.48\textwidth]{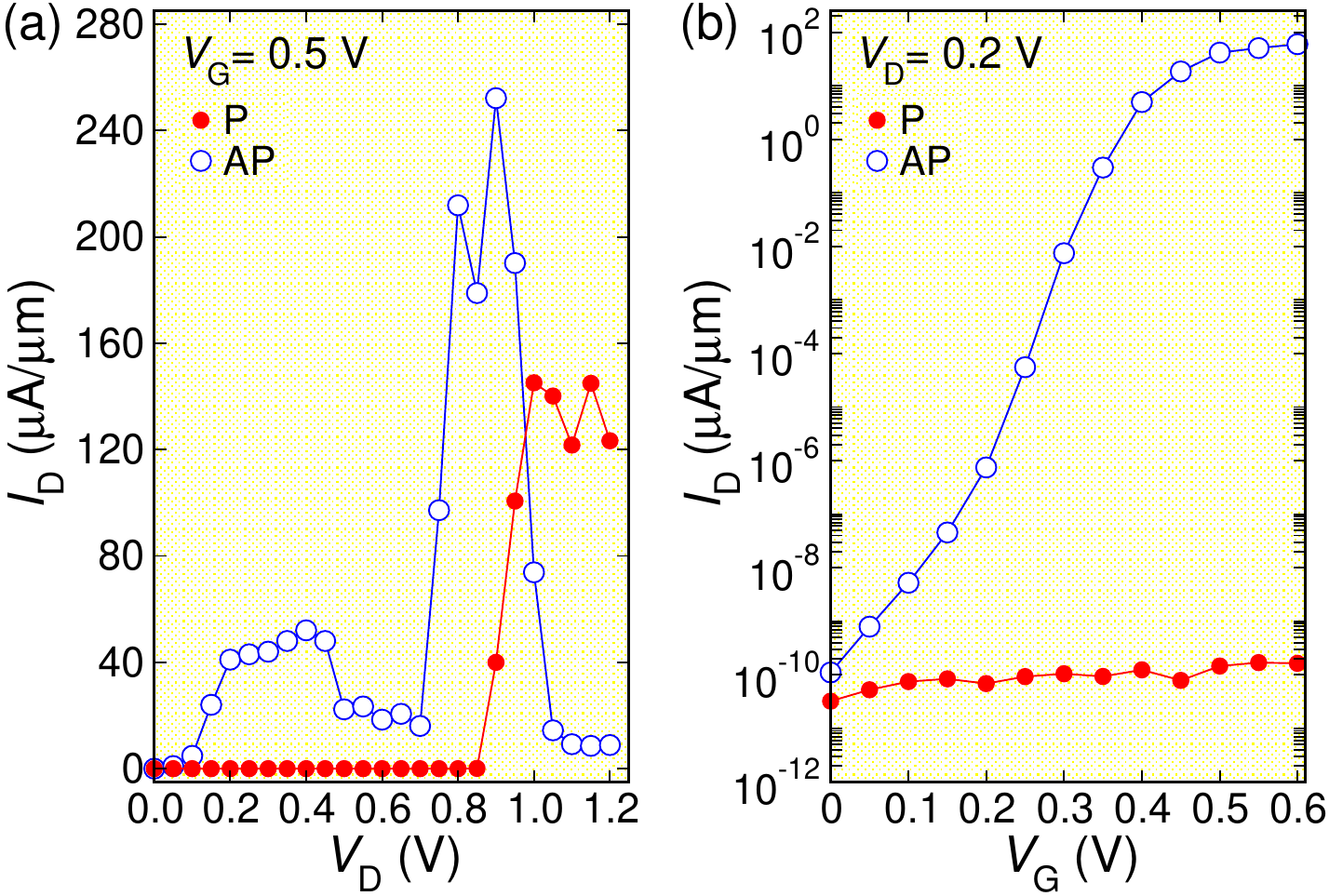}
\vspace{-0.6 cm}
\caption{(a) The output characteristics  ($I_{\mathrm{D}}$-$V_{\mathrm{D}}$)  of a 12 nm channel
length vertical VS$_2$/Ga$_2$O$_2$ heterojunction FET. Drain current ($I_{\mathrm{D}}$) is plotted 
as a function of source-drain bias voltage ($V_{\mathrm{D}}$) for a fixed gate voltage of 0.5\,V, 
comparing parallel (P) and antiparallel (AP) magnetization configurations of the source and drain 
electrodes at 300\,K. (b) Transfer characteristics ($I_{\mathrm{D}}$-$V_{\mathrm{G}}$) of the  same device
at 0\,K. Drain current ($I_{\mathrm{D}}$) is plotted as a function of gate voltage ($V_{\mathrm{G}}$) for a 
fixed source-drain voltage of 0.2\,V, comparing P and AP magnetization configurations. }
\label{fig9}
\end{figure}

Given the potential of SGS materials for ideal switching, we examine the performance of a vertical 
VS$_2$/Ga$_2$O$_2$  heterostructure FET under antiparallel and parallel electrode magnetization 
configurations. As previously discussed, the antiparallel magnetization configuration of the type-II 
SGS source and drain electrodes is crucial for low-voltage, energy-efficient transistor operation. 
In contrast, parallel magnetization configuration results in an off-state for drain-source voltages below 
the drain electrode's external spin gap ($E_{\mathrm{g}}^{\mathrm{E}}$) of approximately 0.8 eV. 
Fig.\,\ref{fig9}(a) illustrates this behavior, showing a sharp increase in drain current ($I_{\mathrm{D}}$) 
only after the drain-source bias voltage ($V_{\mathrm{D}}$) exceeds 0.8 V. This phenomenon is attributed 
to the electronic structure of the source and drain electrodes [Fig.\,\ref{fig6}(b)] and the coherent 
transport mechanism, which prevents spin-up electrons from reaching the drain due to the spin gap. 
For $V_{\mathrm{D}}$ = 1 V and a gate voltage of 0.5 V, $I_{\mathrm{D}}$ surpasses 140 $\mu$V/$\mu$m. 
Conversely, antiparallel magnetization configuration leads to a non-monotonic $I_{\mathrm{D}}$-$V_{\mathrm{D}}$ 
relationship due to the electrode's 2D electronic structure. For example, $I_{\mathrm{D}}$ nearly vanishes 
at  $V_{\mathrm{D}}$ = 1.1 V due to a gap in unoccupied spin-down states [Fig.\,\ref{fig6}(b)]. Overall, 
our calculated transfer and output characteristics align well with the device concept presented earlier.

The distinct behavior of our vertical VS$_2$/Ga$_2$O$_2$  heterostructure FET under parallel and antiparallel 
electrode magnetization configurations manifests as a pronounced non-local GMR effect. As shown in Fig.\,\ref{fig9}(b), 
the drain current ($I_{\mathrm{D}}$) is entirely suppressed for the parallel configuration due to the unique 
band structure of the type-II SGS electrodes, which prohibits current flow. In contrast, the antiparallel 
configuration enables typical transistor operation with $I_{\mathrm{D}}$ exponentially increasing with gate voltage, 
resulting in a 100\%  non-local GMR effect at 0 K. It is important to note that the current version of the 
QuantumATK package does not incorporate the necessary treatment of temperature effects via Fermi-Dirac distribution 
for SGS materials as discussed in Ref.\,\cite{csacsiouglu2020half}. Consequently, the transfer ($I_{\mathrm{D}}$-$V_{\mathrm{G}}$) 
characteristics presented in Fig.\,\ref{fig9}(b) are limited to 0 K. Conversely, for the antiparallel magnetization 
configuration, temperature effects are considered, and the corresponding transfer characteristics are shown in Fig.\,\ref{fig8} at 300 K.

To fully explore the potential of SGS materials for FET applications, a comprehensive investigation encompassing 
a wider range of compounds is necessary. While this study focuses on the vertical VS$_2$-based FET as a proof-of-concept, 
the promising results motivate further exploration of the materials listed in Table\,\ref{table1}. Type-II SGS materials, such 
as VSSe, are anticipated to exhibit similar device characteristics, while type-I SGS compounds like Ti$_4$Cl$_4$Te$_4$ may 
offer comparable functionalities with potentially larger SS values due to electron transport in both spin channels 
(see Fig.\,\ref{fig1}). These materials warrant in-depth theoretical and experimental investigations to assess 
their suitability for high-performance FETs. To further enhance FET performance, we turn our attention to SGMs. Unlike 
SGSs, SGMs offer a broader range of functionalities when integrated as source and drain electrodes in FETs, as previously 
detailed in the device concept section. Essentially, SGMs can be considered as cold metals with an additional degree 
of freedom: spin polarization. This enables not only the conventional steep-slope transistor operation and 
gate-tunable NDR effect associated with cold metals but also introduces the non-local GMR effect. A critical parameter 
influencing the SS value of cold-metal-based FETs is the energy gap edge ($W_{\mathrm{m}}^{\mathrm{E}}$) above the Fermi 
level ([Fig.\,\ref{fig1}(f) and (g)]. Traditional 2D cold metals exhibit relatively large $W_{\mathrm{m}}^{\mathrm{E}}$ 
values, typically between 0.5 eV and 1 eV, which can hinder device performance \cite{sasioglu2023theoretical,wang2021intercalated}. 
In contrast, the SGMs presented in Table II offer significantly smaller $W_{\mathrm{m}}^{\mathrm{E}}$ values, 
ranging from 0.11 eV to 0.75 eV, potentially leading to improved FET characteristics.

Future research should focus on heterostructures comprising different material combinations to fully realize the 
potential of SGS and SGM materials for FET applications. While this study focused on devices with identical source 
and drain electrodes, enabling conventional MOSFET-like operation, the integration of distinct SGS or SGM materials offers 
exciting possibilities. For instance, a FET incorporating a type-II SGS source electrode and a type-IV SGS drain electrode 
could function as a reconfigurable dynamical diode, allowing current flow in a specific direction determined by electrode 
magnetization. This concept, previously proposed by the authors and experimentally demonstrated, holds significant promise 
for novel device functionalities \cite{sasioglu2019proposal,maji2022demonstration}. By systematically investigating various 
material combinations, it is anticipated that a rich landscape of device behaviors can be explored, leading to the 
development of innovative electronic and spintronic devices.

Finally, it is crucial to acknowledge the limitations of the present study in the context of practical device 
implementation. While our theoretical investigations demonstrate the promising potential of SGS- and SGM-based 
FETs, several factors warrant further consideration. Notably, the Curie temperatures of the investigated materials, 
including VS$_2$, are significantly below room temperature. This necessitates the exploration of SGS and SGM 
compounds with higher Curie temperatures for viable device applications. Additionally, the temperature dependence 
of the SGS and SGM characteristics, which has been assumed to be negligible in this study, requires thorough 
investigation. A comprehensive understanding of how these properties evolve with temperature is essential for 
accurate device modeling and optimization. Future research should focus on identifying and characterizing SGS 
and SGM materials with high Curie temperatures and exploring the impact of temperature on their electronic
and magnetic properties to pave the way for the realization of high-performance spintronic FETs.

\section{Summary and Conclusions}

This work presents a comprehensive exploration of the potential of SGSs and SGMs for realizing innovative spintronic 
transistors. By harnessing the unique spin-dependent transport properties of these materials, we propose a novel device 
concept that significantly advances beyond conventional semiconductor technology. The integration of SGSs and/or SGMs 
into FET architectures offers a promising pathway to overcome the longstanding subthreshold swing limitations, 
enabling low-power and high-performance operation.  We perform a comprehensive screening of the computational 
2D materials database to identify suitable SGS and SGM materials for the proposed devices. For device simulations, 
we select VS$_2$ as the SGS material. Our theoretical investigations, focusing on a vertical VS$_2$/Ga$_2$O$_2$/VS$_2$ 
heterostructure FET, demonstrates the feasibility of achieving remarkably low SS values, high on/off ratios, and 
significant non-local GMR effects. These findings underscore the potential of SGS-based FETs for revolutionizing 
electronic and spintronic devices. Furthermore, the incorporation of SGMs introduces additional functionalities, 
such as NDR effect, expanding the device capabilities and enabling the exploration of novel circuit concepts.

To fully realize the potential of SGS and SGM materials in FET applications, several critical challenges must be 
addressed. While our study provides a foundational understanding, a comprehensive exploration of the material 
space is essential to identify SGS and SGM compounds with optimal properties, including high Curie temperatures. 
Understanding the temperature dependence of these materials' electronic and magnetic characteristics is crucial 
for accurate device modeling and practical implementation. Additionally, the development of advanced fabrication 
techniques and device engineering strategies is necessary to bridge the gap between theoretical concepts and 
functional devices. Despite these challenges, the integration of SGS and SGM-based FETs into electronic systems 
holds the promise of transformative advancements. By combining the advantages of low-power operation, high performance, 
and novel functionalities, these devices can pave the way for innovative computing architectures and applications. 
The potential impact extends beyond traditional electronics, with implications for fields such as neuromorphic computing,
multi-valued logic, logic-in-memory computing, and artificial intelligence.

\begin{acknowledgments}
This work was supported by SFB CRC/TRR 227 of Deutsche Forschungsgemeinschaft (DFG) and by the 
European Union (EFRE) via Grant No: ZS/2016/06/79307. 
\end{acknowledgments}


%

\pagebreak
\onecolumngrid

\section*{Supplemental Material\\[1cm]
Multifunctional steep-slope spintronic transistors with \\ spin-gapless-semiconductor or spin-gapped-metal electrodes}

\subsection*{Supplementary Figures}

The supplementary figures provide additional insights into the structural, electronic, and device properties of the 
vertical VS\(_2\)/Ga\(_2\)O\(_2\) heterostructure FET. Figure S1 and Figure S2 present the crystal structures
of the 2D spin gapless semiconductors (listed in Table I) and 2D spin gapped metals (listed in Table II), respectively.
Arrows indicate the magnetic moments, with their sizes proportional to their respective magnitudes.
Figure S3 presents the calculated projected band structure, revealing the energy band alignment and potential for bandgap 
engineering. Figure S4 depicts the device density of states (DDOS) for both majority and minority spin electrons under 
flat-band conditions, providing information on the distribution of electronic states within the device. Figure S5 
focuses on the DDOS for minority spin electrons under different device states (off-state and on-state), allowing for 
an analysis of how the electronic structure changes with device operation.

\begin{figure}[!ht]
  \centering
   \includegraphics[width=0.99\textwidth]{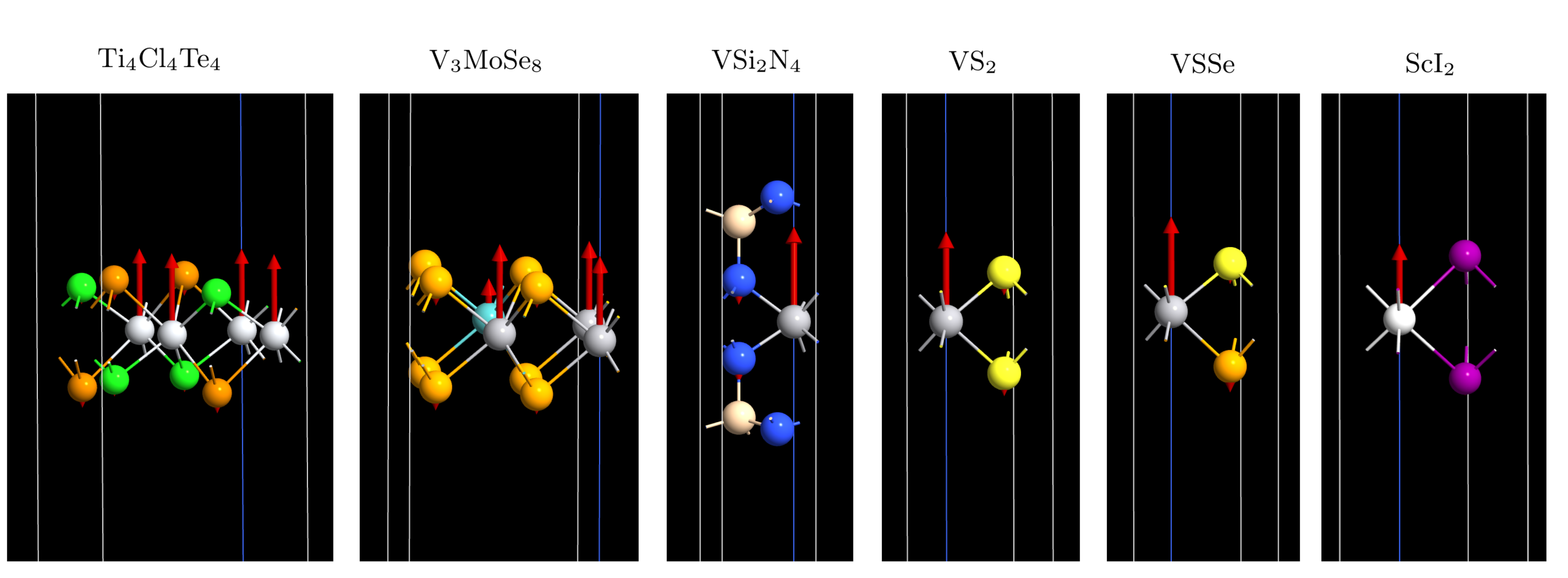}
   \vspace*{-0.1cm}
   \caption{Crystal structures of the 2D spin gapless semiconductors listed in Table I of the main text. Arrows indicate the magnetic moments, with their sizes proportional to the respective magnitudes. For better visibility, sublattice magnetic moments in all compounds are scaled up by a factor of two.}
   \label{fig:p_bands}
\end{figure}

\begin{figure}[!ht]
    \centering
    \includegraphics[width=0.99\textwidth]{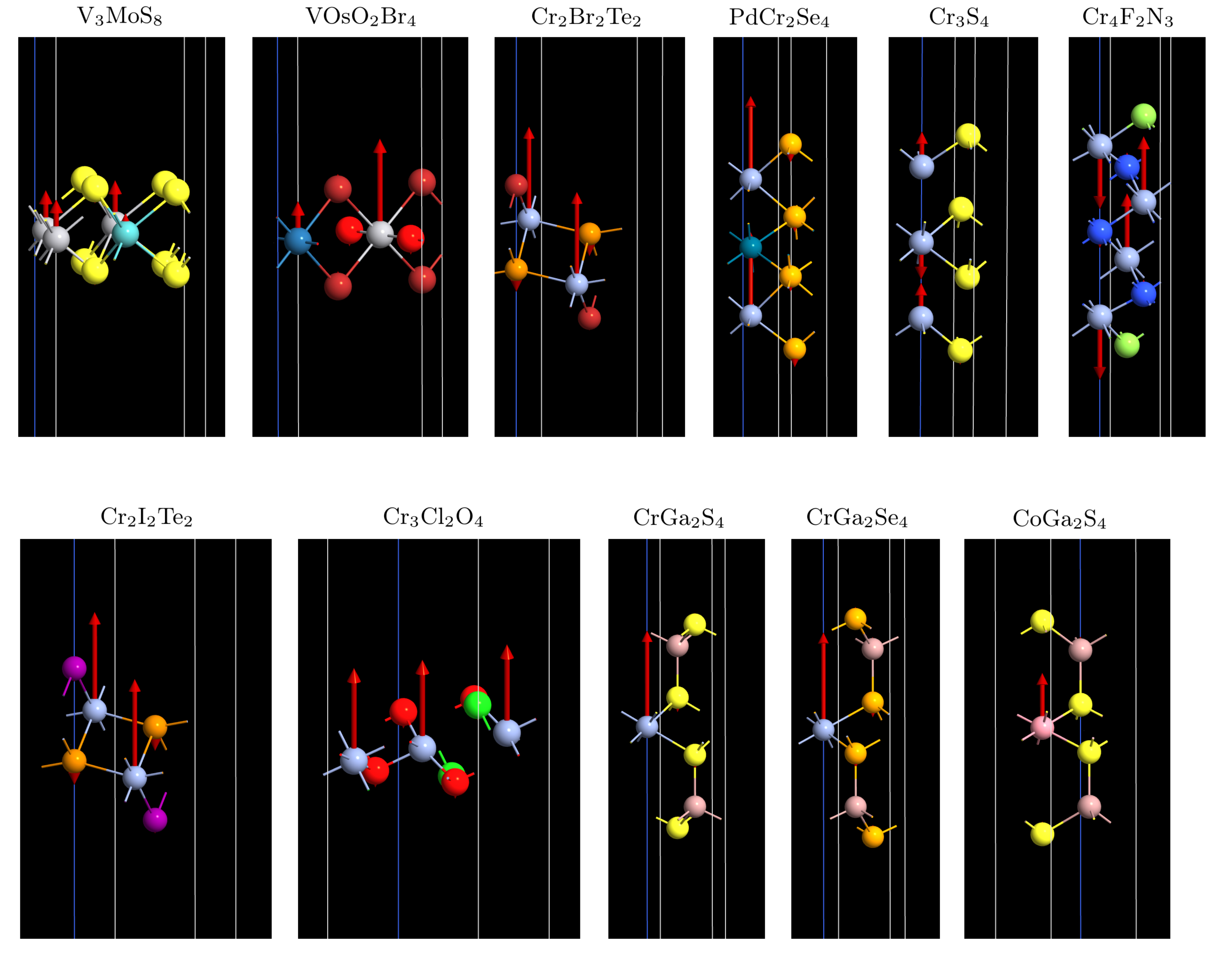}
    \vspace*{-0.3cm}
    \caption{Crystal structures of the 2D spin-gapped metals listed in Table II of the main text. Arrows indicate the magnetic moments, with their sizes proportional to their respective magnitudes. For better visibility, sublattice magnetic moments are scaled up by a factor of two in V- and C-based compounds, while in Cr$_3$S$_3$ and Cr$_4$F$_2$N$_3$ they are scaled down by a factor of 0.3 and 0.7, respectively.}
    \label{fig:p_bands}
\end{figure}

\begin{figure}[!ht]
    \centering
    \includegraphics[width=0.8\textwidth]{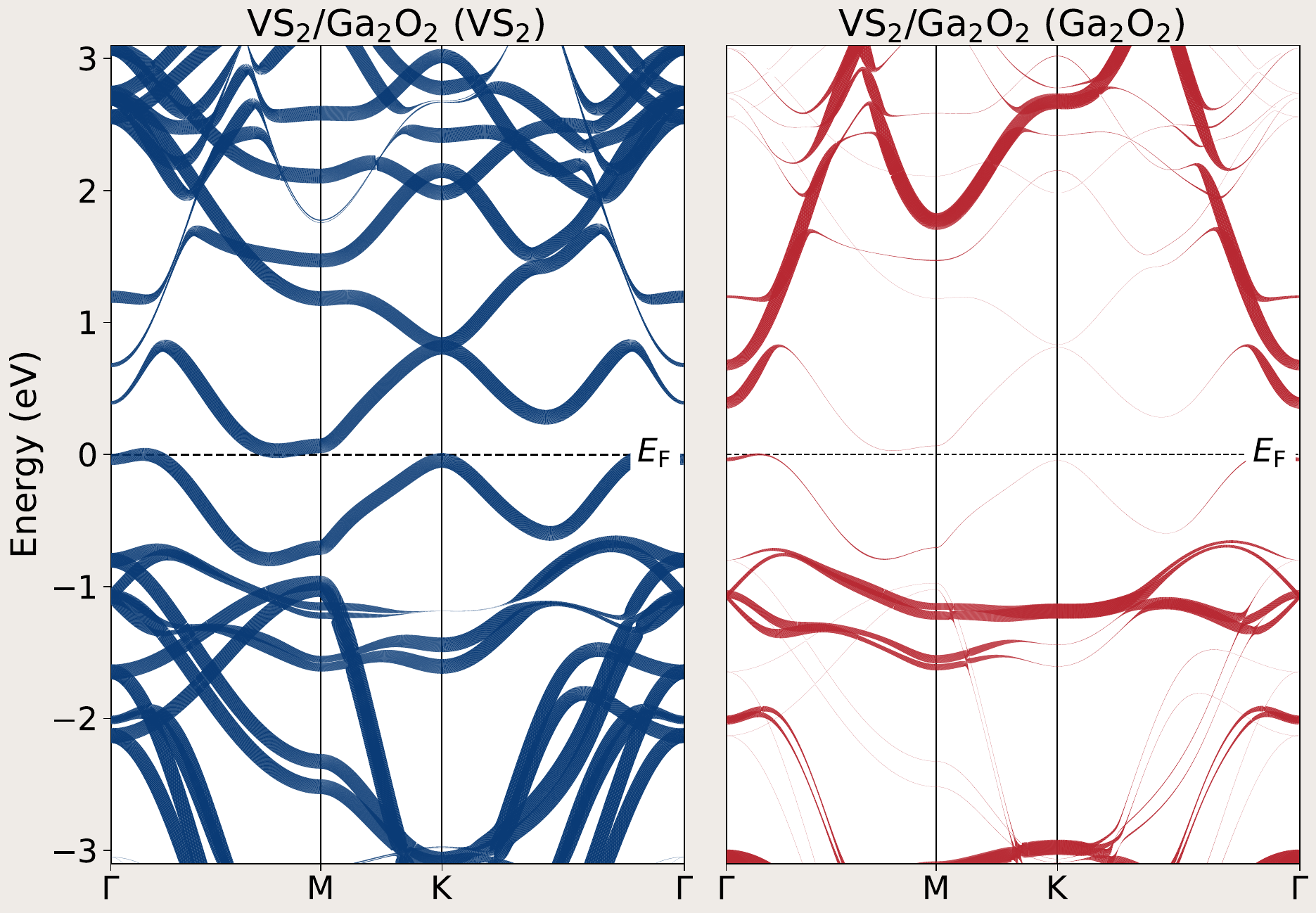}
    \vspace*{-0.01cm}
    \caption{Calculated projected band structure of the VS$_2$/Ga$_2$O$_2$  heterostructure along high-symmetry directions. 
    Left panel: Projection onto the VS$_2$ layer. Right panel: Projection onto the Ga$_2$O$_2$ layer.  The dashed black 
    line indicates the Fermi level, which is set to zero energy.}
    \label{fig:p_bands}
\end{figure}

\begin{figure}[t]
    \centering
    \includegraphics[width=0.8\textwidth]{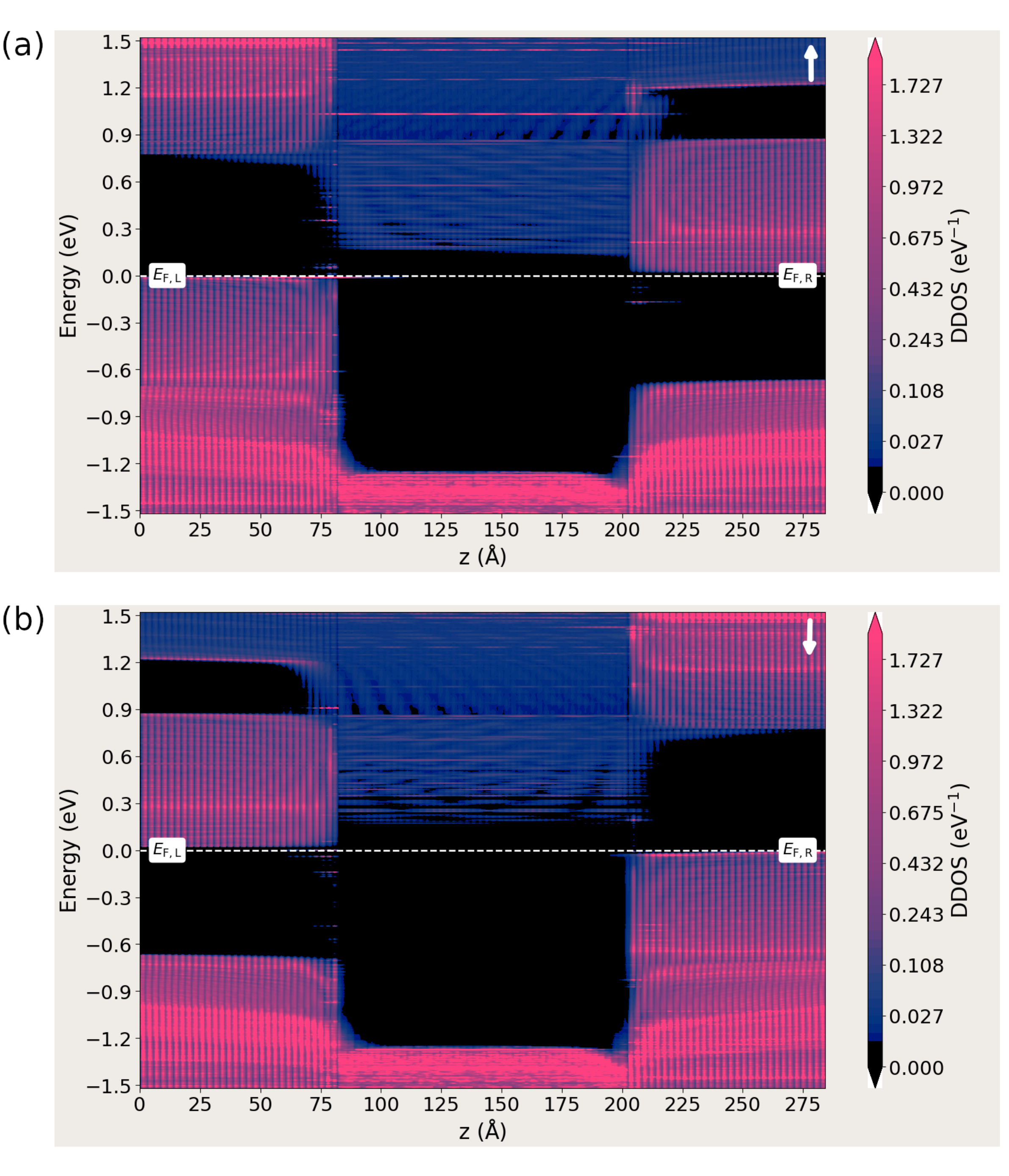}
    \vspace*{-0.3cm}
    \caption{Projected device density of states (DDOS) for (a) majority and (b) minority spin electrons in a 12 nm channel length 
    vertical VS$_2$/Ga$_2$O$_2$  heterostructure FET under flat-band conditions (zero gate and drain bias voltages). White dashed 
    lines indicate the Fermi level of the source and drain electrodes, respectively.}
    \label{fig:PDOS_1}
\end{figure}

\begin{figure}[t]
    \centering
    \includegraphics[width=0.8\textwidth]{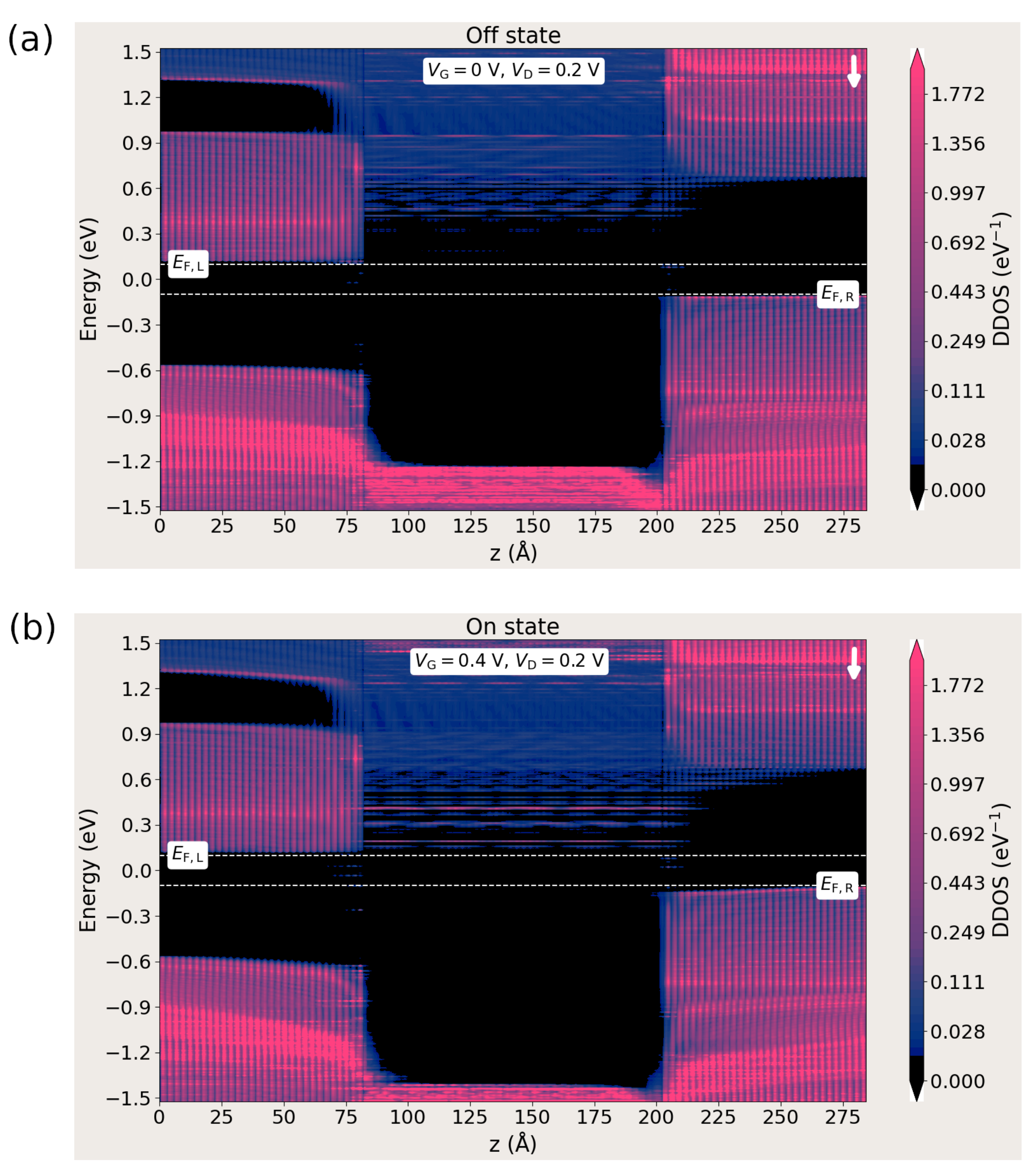}
    \vspace*{-0.1cm}
    \caption{Projected device density of states (DDOS) for minority spin electrons in a 12 nm channel length 
    vertical VS$_2$/Ga$_2$O$_2$  heterostructure FET  (a) for off-state and (b) for on-state. Upper and lower 
    white dashed lines indicate the Fermi levels of the source and drain electrodes, respectively. }
    \label{fig:PLDOS}
\end{figure}

\end{document}